\documentclass[prl,twocolumn,showpacs]{revtex4}
\usepackage{amsmath,amssymb,graphicx,color,bm,soul}
\usepackage{latexsym}
\usepackage{dcolumn}
\usepackage{epsf}
\usepackage{float}

\usepackage{hyperref}
\hypersetup{
    colorlinks=true,
    linkcolor=black,%blue,
    filecolor=magenta,      
    urlcolor=cyan,
}
 
\begin{document}

\title{Continuous wave single photon transistor based on a superconducting circuit}

\author{Oleksandr Kyriienko}
\affiliation{The Niels Bohr Institute, University of Copenhagen, Blegdamsvej 17, DK-2100 Copenhagen, Denmark}

\author{Anders S. S{\o}rensen}
\affiliation{The Niels Bohr Institute, University of Copenhagen, Blegdamsvej 17, DK-2100 Copenhagen, Denmark}

\date{\today}

\begin{abstract}
We propose a microwave frequency single photon transistor which can operate under continuous wave probing, and represents an efficient single microwave photon detector. It can be realized using an impedance matched system of a three level artificial ladder-type atom coupled to two microwave cavities connected to input/output waveguides. Using a classical drive on the upper transition, we find parameter space where a single photon control pulse incident on one of cavities can be fully absorbed into hybridized excited states. This subsequently leads to series of quantum jumps in the upper manifold and the appearance of a photon flux leaving the second cavity through a separate input/output port. The proposal does not require time variation of the probe signals, thus corresponding to a passive version of single photon transistor. The resulting device is robust to qubit dephasing processes, possesses low dark count rate for large anharmonicity, and can be readily implemented using current technology. 
\end{abstract}

% 42.50.-p Quantum optics
% 42.50.Lc Quantum fluctuations, quantum noise, and quantum jumps
% 85.25.-j Superconducting devices
% 03.67.Lx Quantum computation architectures and implementations

% 85.25.Pb Superconducting infrared, submillimeter and millimeter wave detectors
% 42.50.Nn Quantum optical phenomena in absorbing, amplifying, dispersive and conducting media; cooperative phenomena in quantum optical systems

\maketitle

% 71.36.+c 	Polaritons (including photon-phonon and photon-magnon interactions)
% 42.50.Wk	Mechanical effects of light on material media, microstructures and particles
% 07.10.Cm	Micromechanical devices and systems
% 42.65.Pc	Optical bistability, multistability, and switching, including local field effects
% 71.35.-y 	Excitons and related phenomena
% 78.67.-n  Optical properties of low-dimensional, mesoscopic, and nanoscale materials and structures
% 42.65.-k	Nonlinear optics
% 85.85.+j	Micro- and nano-electromechanical systems (MEMS/NEMS) and devices
% 42.50.Lc	Quantum fluctuations, quantum noise, and quantum jumps
% 42.79.Hp 	Optical processors, correlators, and modulators
% 42.50.-p - Quantum optics
% 71.36.+c - Polaritons
% 42.50.Dv - Antibunched photon states
% 42.50.Ar - Photon statistics and coherence theory

Electronic transistors---devices where weak electrical signal controls a strong probe from a source---lie at the heart of modern electronics, and has led to vast development of classical computing devices. By analogy, in the realm of quantum computing a similar device was contrived, where a single photon control pulse triggers the transmission of a strong coherent probe, and was named a \textit{single photon transistor} (SPT) \cite{Chang2007}. The operation of the SPT device proposed in Ref. \cite{Chang2007} is based on a time-dependent control of the drive and a strong atom-photon interaction. In the optical domain single photon transistors along these lines were realized with neutral atoms embedded in an optical cavity \cite{Chen2013,Tiecke2014,Reiserer2014}, a quantum dot in a waveguide \cite{Kim2013,Javadi2015}, or an ultracold gas with Rydberg interactions \cite{Tiarks2014,Gorniaczyk2014}. So far the achieved efficiencies have been limited, but if this is improved, a single photon transistor could represent a powerful tool for coherent state manipulation and quantum information processing \cite{Duan2004}. Importantly, a SPT can also serve as efficient single photon detector (SPD), as it amplifies a single photon signal by a large gain.

Recently, a microwave frequency range counterpart of quantum optics---circuit quantum electrodynamics (cQED) \cite{Wallraff2004}---has emerged as a highly promising platform for quantum computation \cite{Barends2014,Kelly2015,Riste2015,Reed2012}. Based on high-quality superconducting microwave cavities combined with Josephson junction-based artificial atoms, it enables a strong light-matter coupling even at the single photon level, and allows studying numerous nonlinear microwave quantum optics phenomena \cite{Bishop2009,Houck2007,Hoi2013,Pechal2014}. The development of a simple and efficient single microwave photon detector is still an open question \cite{Sathyamoorthy2015}. The suggested realizations include SPDs based on current biased Josephson junctions \cite{Romero2009}, catching an inverted time-controlled pulse \cite{Yin2013,Wenner2014}, transmon chain linked with non-reciprocal elements \cite{Sathyamoorthy2014}, and double quantum dot structures \cite{Wong2015}. Also, several schemes for cQED-based single photon transistors have been proposed \cite{Neumeier2013,Manzoni2014}. Ultimately, however these proposals rely on active time-control of the system and input single photon pulse, which complicates the detection process and limits the applicability. Lately such a time dependent protocol based on an impedance artificial $\Lambda$ atoms was proposed \cite{Koshino2015} and experimentally realized \cite{Inomata2016}. This protocol can be extended to perform time-independent detection \cite{Koshino2016} if highly anharmonic systems with long coherence time can be constructed.

\begin{figure}
\includegraphics[width=1.0\linewidth]{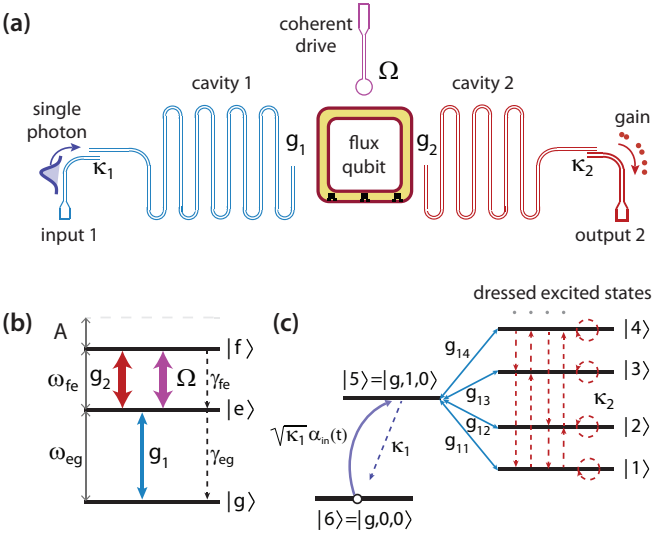}
\caption{(color online). (a) Sketch of the system, showing a driven qubit coupled to two superconducting cavities, each connected to a separate input/output line. (b) Energy diagram of three lowest artificial atom levels with associated cavity and drive couplings. (c) Relevant manifold of SPT operational levels with dressed excites states.}
\label{fig:sketch}
\end{figure}

Here, we propose a single photon transistor which can operate under continuous wave (cw) probe conditions, where a single photon control pulse triggers an avalanche of gain photons. It represents a passive device which does not require signal and probe timing, largely extending its applicability. The proposed device is robust to imperfections and is particularly insensitive to qubit dephasing. The generic idea relies on a three level ladder atom [Fig. \ref{fig:sketch}(b)], with its lower transition weakly coupled to the first input cavity, and the upper transition strongly driven by a classical source as well as strongly coupled to a second output cavity. A single photon entering the input port transfers the atom to the excited subspace through an impedance matching mechanism similar to Refs. \cite{Koshino2015,Inomata2016,Koshino2016,Inomata2014,Koshino2013,Koshino2010}. In the excited subspace, a number of quantum jumps between dressed atom-cavity states leads to an enhanced output signal.

\textit{System and Hamiltonian.---}As a particular realization of cw microwave SPT we propose a superconducting artificial atom with three states $|g\rangle$, $|e\rangle$, and $|f\rangle$, which is coupled to two separate microwave cavities (modes $\hat{a}_1$ and $\hat{a}_2$), both connected to input-output waveguide channels with coupling constants $\kappa_1$ and $\kappa_2$ [see sketch in Fig. \ref{fig:sketch}(a)]. The qubits should have versatile connectivity \cite{Riste2015,Barends2013,Chen2014} and sizeable anharmonicity, making flux \cite{Chiorescu2004} and fluxonium \cite{Manucharyan2009} qubits desirable. Its lower $|g\rangle$-$|e\rangle$ transition is resonantly coupled to a cavity mode $\hat{a}_1$ with a perturbative coupling $g_1$, while the upper $|e\rangle$-$|f\rangle$ transition is strongly coupled to cavity $\hat{a}_2$ with strength $g_2$ [Fig. \ref{fig:sketch}(b)]. Additionally, the upper levels are driven by a classical qubit drive of strength $\Omega$.
%%%
%%%

The Hamiltonian of the system reads
%%%
\begin{equation}
\label{eq:H}
\hat{\mathcal{H}} = \hat{\mathcal{H}}_{\mathrm{sys}} + \hat{\mathcal{H}}_{\mathrm{wgd,1}} + \hat{\mathcal{H}}_{\mathrm{wgd,2}},
\end{equation}
%%%
where $\hat{\mathcal{H}}_{\mathrm{sys}}$ corresponds to the system Hamiltonian written in the rotating frame as [\onlinecite{SM}a]
\begin{align}
\label{eq:Hsys}
&\hat{\mathcal{H}}_{\mathrm{sys}} = \delta_e \sigma_{\mathrm{ee}} + (\delta_e + \delta_f) \sigma_{\mathrm{ff}} + \delta_{\mathrm{cav},1} \hat{a}_1^\dagger \hat{a}_1 + \delta_{\mathrm{cav},2} \hat{a}_2^\dagger \hat{a}_2 \\ \notag &+ g_1 (\hat{a}_1^\dagger \sigma_{\mathrm{eg}}^- + \sigma_{\mathrm{eg}}^+ \hat{a}_1) + g_2 (\hat{a}_2^\dagger \sigma_{\mathrm{fe}}^- + \sigma_{\mathrm{fe}}^+ \hat{a}_2) + \Omega (\sigma_{\mathrm{fe}}^- + \sigma_{\mathrm{fe}}^+),
\end{align}
with detunings $\delta_e = \omega_{\mathrm{eg}} - \omega_s$, $\delta_f = \omega_{\mathrm{fe}} - \omega_d$, $\delta_{\mathrm{cav},1} = \omega_{\mathrm{cav},1} - \omega_s$, and $\delta_{\mathrm{cav},2} = \omega_{\mathrm{cav},2} - \omega_d$. Here $\omega_{\mathrm{eg}}$, $\omega_{\mathrm{fe}}$, $\omega_{\mathrm{cav},1}$, $\omega_{\mathrm{cav},2}$ denote energy separations between qutrit levels and energies of microwave cavities, sequentially ($\hbar = 1$). $\omega_s$ is the single input photon central frequency and $\omega_d$ is the frequency of the classical drive. $\sigma_{\mathrm{mn}}^+ = |m \rangle\langle n|$ ($\sigma_{\mathrm{mn}}^- = |n \rangle\langle m|$) denotes qubit raising (lowering) operator, and $\sigma_{\mathrm{mm}} = |m \rangle\langle m|$. $\hat{\mathcal{H}}_{\mathrm{wgd,j}}$ describes the coupling to the waveguides $j=1,2$ [\onlinecite{SM}a].
%, given by
%
%\begin{equation}
%\label{eq:Hwgd}
%\hat{\mathcal{H}}_{\mathrm{wgd,j}} = \int_{-\infty}^{+\infty} dp \omega_{j,p} \hat{b}_{j,p}^\dagger \hat{b}_{j,p} - i \int_{-\infty}^{+\infty} dp \sqrt{\frac{\kappa_j}{2\pi}} \left( \hat{a}_j^\dagger \hat{b}_{j,p} - H.c. \right),
%\end{equation}
%
%where $\hat{b}_{i,p}^\dagger$ ($\hat{b}_{i,p}$) denotes creation (annihilation) operator for waveguide mode $i=1,2$ with 1D wavevector $p$. 
The expression for $\hat{\mathcal{H}}_{\mathrm{sys}}$ assumes infinitely large anharmonicity, which precludes parasitic couplings to other-than-resonant qubit transitions, and we assume long decay/dephasing times for the qubit. These assumptions will be revisited later. Also, in the following we consider zero detuning for the incoming single photon pulse, $\delta_e = \delta_{\mathrm{cav},1} = 0$, a resonant microwave drive, $\delta_f = 0$, and a cavity resonant to the upper transition, $\delta_{\mathrm{cav},2} = 0$.

\textit{Operational principle.---}First, a single photon pulse enters through the input channel, which is loaded by the joint qubit-cavity system with Hilbert space $\{|g\rangle,|e\rangle,|f\rangle \}\otimes|n_1\rangle\otimes|n_2\rangle \equiv |m,n_1,n_2\rangle$ ($m=g,e,f$). In the input stage we consider a weak excitation such that $n_1$ is restricted to vacuum or a single excitation. Then, the relevant subspace of states contains the ground state $|g,0,0\rangle$, the first cavity excited state $|g,1,0\rangle$, and the subspace $\{|e,0,n_2\rangle,|f,0,n_2\rangle\}$ which we call excited states. Due to the strong couplings $g_2$ and $\Omega$, the excited states become hybridized, and it is convenient to introduce dressed metastable states. For the lowest cavity 2 occupation with $n_2 =0,1$ this embeds a subspace $\mathcal{M}=\{|1\rangle,|2\rangle,|3\rangle,|4\rangle\}$ (Fig. \ref{fig:sketch}(c) and [\onlinecite{SM}b]), and higher states can be included analogously. Due to the admixture of the $|e,0,0\rangle$ level, each dressed state is coupled to $|g,1,0\rangle$ with a modified constant $g_{1m}$. Considering $g_1$ to be perturbative, the coupling of the first cavity to the dressed states $\mathcal{M}$ works as an effective decay channel. By controlling the drive strength $\Omega$ and coupling parameters, this effective total decay rate $\Gamma_{\mathrm{set}}$ can be made equal to the coupling of the cavity to the first waveguide $\kappa_1$, reducing the system to an impedance matched $\Lambda$ system attached to a single-sided waveguide [\onlinecite{SM}c]. Such impedance matching (IM) has already proven to be useful for cQED circuits, leading to photo detection \cite{Inomata2016}, as well as proposals for microwave downconversion \cite{Inomata2014,Koshino2013} and gates for flying qubits \cite{Koshino2010}. This allows for a full absorption of a single photon pulse by the metastable states, leading to near-unity single photon switching and photon detection without the need for temporally varying control fields.

Once the setting stage to the excited states manifold is completed, the second decay channel $\kappa_2$ leads to quantum jumps between the dressed states [Fig. \ref{fig:sketch}(a), red dashed lines], where a series of jumps within the full $\{|e,0,n_2\rangle,|f,0,n_2\rangle\}$ ($n_2 = 0,1,..,\mathcal{N}_2$) subspace occur. Radiative jumps between the states lead to a flux leaving cavity 2 through the output port, resulting in an enhancement of the signal. Once the system recovers to the ground state $|g,0,0\rangle$, the SPT duty cycle is finalized. The SPT can thus serve as a highly efficient single photon detector, and corresponds to a microwave version of a single photon avalanche diode.
%%%
\begin{figure}
\includegraphics[width=1.0\linewidth]{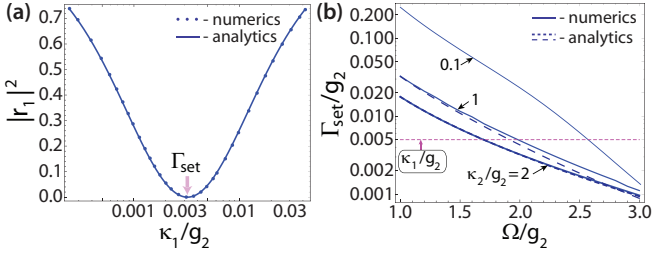}
\caption{(color online). (a) Reflection coefficient for the first cavity as a function of input coupling rate $\kappa_1$ (log scale). The dotted line is a full numerical calculation, and the solid line represents  Eq. (\ref{eq:reflection}) with $\Gamma_{\mathrm{set}}$ given by Eq. (\ref{eq:Gamma_set}). Here $\Omega/g_2 = 2$, $g_1/g_2 = 0.05$, and $\kappa_2/g_2 = 2$. (b) Tunability of the effective setting rate as a function of qubit drive strength $\Omega$ and varying $\kappa_2$ with $g_1/g_2 = 0.05$. The full lines are the result of a numerical simulation which agree with the analytical prediction of Eq. (4) for $\kappa_2 \gtrsim 1$. The dashed horizontal line represent the impedance matching condition which can always be obtained by varying $\Omega$.}
\label{fig:input}
\end{figure}
%%%

\textit{Input: impedance matching.---}First, we characterize the single photon input stage. To define the IM condition analytically we exploit the effective operator theory \cite{Reiter2012} to estimate the decay rate $\Gamma_{\mathrm{set}}$ from $|g,1,0\rangle$ to the metastable states [\onlinecite{SM}d]. Considering $g_1 < g_2, \Omega$ and up to $\mathcal{N}_2 = 2$ photons, we find
\begin{equation}
\label{eq:Gamma_set}
\Gamma_{\mathrm{set}} = \frac{16 \kappa_2 g_1^2 g_2^2 (16 g_2^2 + 4 \kappa_2^2 + \Omega^2)}{4 \kappa_2^2 (4 g_2^2 + \kappa_2^2) \Omega^2 + 5 \kappa_2^2 \Omega^4 + \Omega^6},
\end{equation}
which provides a good estimate for $\kappa_2 / g_2 \gtrsim 1.5$, where the relevant processes happen within the lowest Fock states. The setting rate for smaller $\kappa_2 / g_2$ can be derived by increasing $\mathcal{N}_2$ [\onlinecite{SM}d].

To test the IM condition numerically, we exploit the input-output theory for the Hamiltonian (\ref{eq:H}) by deriving the Heisenberg equations of motion for the system operators, and assuming a weak coherent input. The efficiency of the setting stage is quantified by numerically calculating the reflection coefficient $|r_1|^2 = |\langle \hat{a}_{\mathrm{out},1}\rangle/\langle \hat{a}_{\mathrm{in},1}\rangle|^2$ at the input port. In the simulation a reflection minimum is obtained at the impedance matched setting rate $\kappa_1=\Gamma_{\mathrm{set}}$ [Fig. \ref{fig:input}(a)]. The dotted curve corresponds to the full Heisenberg equation calculation, and the solid curve shows the analytical solution for a reflection coefficient of a waveguide coupled to a $\Lambda$ system [\onlinecite{SM}c,~\onlinecite{Pinotsi2008}]
\begin{equation}
\label{eq:reflection}
|r_1|^2 = \frac{(\Gamma_{\mathrm{set}}/\kappa_1 - 1)^2}{(\Gamma_{\mathrm{set}}/\kappa_1 + 1)^2},
\end{equation}
with $\Gamma_{\mathrm{set}}$ provided by Eq. (\ref{eq:Gamma_set}). 
%We note that by tuning the system parameters $g_1$, $g_2$, $\Omega$, and $\kappa_2$ IM conditions can always be met. 
%Here and further, we consider the coupling $g_2$ to be relevant energy scale in the system, and use it to perform dimensionless analysis.
%
The dependence of the setting rate on the classical microwave drive strength $\Omega$ allows fine tuning the IM condition. In Fig. \ref{fig:input}(b) we show the dependence of $\Gamma_{\mathrm{set}}/g_2$ on $\Omega$, fixing $\kappa_1 = 0.005 g_2$ and $g_1 = 0.05 g_2$. Plotting the setting rate for three different values of $\kappa_2$ we can find a value of $\Omega/g_2$ for which IM holds.
%%%
\begin{figure}[t!]
\includegraphics[width=1.0\linewidth]{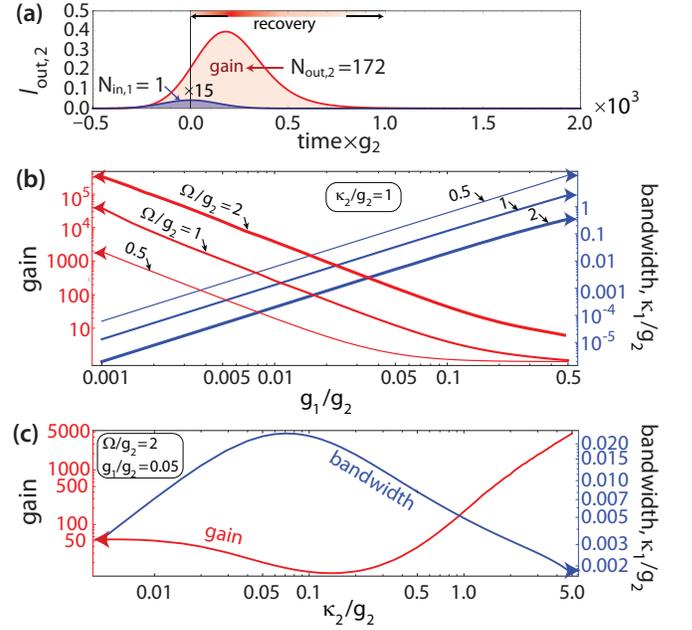}
\caption{(color online). (a) Time dependence of second cavity output calculated for Gaussian single photon pulse incident on the first cavity; $g_1/g_2=0.05$, $\Omega/g_2=2$, $\kappa_2/g_2=1$. (b) SPT gain and bandwidth plotted as function of the first cavity coupling $g_1$ for various drive strengths; $\kappa_1 = \Gamma_{\mathrm{set}}$, $\kappa_2/g_2=1$. (c) SPT gain and bandwidth for varying output coupling $\kappa_2$; $g_1/g_2=0.05$.}
\label{fig:output}
\end{figure}
%%%

\textit{Output: gain of SPT.---}To describe the gain of the transistor, we exploit the Heisenberg equations of motion derived using the input-output relations for two coupling channels combined with a Gaussian-shaped single photon pulse [\onlinecite{SM}e]. For temporal widths of the pulse being larger than $\tau > \kappa_1^{-1}$ and IM arranged, an incoming single photon excitation is fully transferred to the excited states of the qutrit which are highly mixed with the mode of cavity 2. Averaging the Heisenberg equation with the wave function corresponding to the single photon input $|\Psi_{\mathrm{in}}\rangle$, we can extract the intensity of outgoing photons in the second waveguide, $\langle \hat{a}_{\mathrm{out},2}^\dagger \hat{a}_{\mathrm{out},2}\rangle = \kappa_2 \langle \hat{a}_{2}^\dagger \hat{a}_{2}\rangle \equiv I_{\mathrm{out},2}$.
In Fig. \ref{fig:output}(a) we plot the temporal dependence of input 1 and output 2 photon numbers. For a long Gaussian pulse containing a single photon [$N_{\mathrm{in},1} = \int dt I_{\mathrm{in},1}(t) = 1$], we can on average get $N_{\mathrm{out},2}= \int dt I_{\mathrm{out},2}$ photons at the output of the second cavity before the system recovers to the ground state. This represents the \emph{gain} of the single photon transistor, which describes the effective amplification of the single photon signal. The full counting statistics can be obtained by changing to the wave-function Monte-Carlo approach. This confirms that a single input photon leads to numerous emitted photons [\onlinecite{SM}f].

To investigate the gain we consider the input stage to be completed, setting $|\Psi_{\mathrm{start}}\rangle = |e,0,0\rangle$, and calculate the occupation of cavity 2 using a density matrix approach, truncating the cavity Fock space at $\mathcal{N}_2 = 10$ excitations. The gain is highly sensitive to the system parameters: cavity 1 to SQ coupling $g_1$, microwave drive strength $\Omega$, and output coupling $\kappa_2$. Additionally, these parameters set the optimal (IM) value for the input coupling $\kappa_1$, and thereby the \emph{bandwidth} of the single photon detector. In Fig. \ref{fig:output}(b) we show the gain and bandwidth as a function of the coupling $g_1$ for varying $\Omega/g_2$. The plot shows a fast increase of the bandwidth with the coupling constant $g_1$, as it sets the rate at which single photons can get to the excited subspace. Reciprocally, this corresponds to a growth of the recovery rate and a reduction of the gain. Thus, there is a trade-off between gain and bandwidth, indicating that a medium $g_1/g_2$ ratio is favoured for optimal detection.

The gain and bandwidth dependence as a function of $\kappa_2$ is shown in Fig. \ref{fig:output}(c). Choosing $g_1/g_2 = 0.05$ and $\Omega/g_2 = 2$ in order to get high input bandwidth, we find that both gain and bandwidth are non-monotonous functions of $\kappa_2$. In particular, we are interested in the region of $0.5<\kappa_2/g_2<3$, where the gain increases rapidly, while $\kappa_1$ is relatively large.
%%%
\begin{figure}
\includegraphics[width=1.\linewidth]{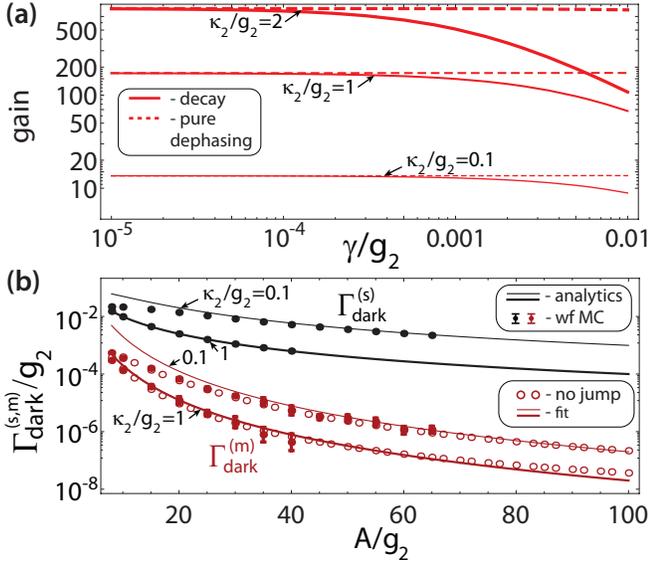}
\caption{(color online). (a) Gain of SPT as a function of qubit decay (solid curves) and pure dephasing (dashed); $\Omega/g_2 = 2$, $g_1/g_2 = 0.05$, $\kappa_1 = \Gamma_{\mathrm{set}}$. (b) The SPT enhanced $\Gamma^{\mathrm{(m)}}_{\mathrm{dark}}$ (lower two curves) and single $\Gamma^{\mathrm{(s)}}_{\mathrm{dark}}$ (upper two curves) dark count rates as a function of the qutrit anharmonicity $A$. Here, $\Omega/g_2 = 2$, $g_1/g_2 = 0.05$. Numerical results were obtained using the wave function Monte-Carlo approach and studying no-jump evolution of the system [\onlinecite{SM}h]. We show analytical results for a single dark count rate derived using effective operator formalism [Eq. (6), black solid curves]. For $\Gamma^{\mathrm{(m)}}_{\mathrm{dark}}$ we also show a fit $\eta \Gamma_{\mathrm{dark,ss}}^{\mathrm{(m)}}$ with $\eta \approx 4$.}
\label{fig:anharmonicity}
\end{figure}
%%%

\textit{Imperfections.---}We now turn to imperfections of the scheme. First, we plot the SPT gain as a function of qubit decay (solid) and pure dephasing (dashed) rate $\gamma$ [\onlinecite{SM}g] for three values of $\kappa_2/g_2$ [see Fig. \ref{fig:anharmonicity}(a)]. While strong decay naturally tends to decrease the gain by introducing an extra non-radiative recovery channel, pure dephasing does not influence the gain, thus removing the need for long qubit coherence time, as opposed to e.g. Refs. \cite{Manzoni2014,Koshino2015}. We also find that even for larger qubit decay the IM conditions can be always satisfied.

Second, we account for a finite qubit anharmonicity $A$ and introduce the residual qutrit-cavity couplings defined by the Hamiltonian [\onlinecite{SM}h]
\begin{equation}
\label{eq:Hres}
\hat{\mathcal{H}}_{\mathrm{res}} = \left(\sqrt{2} g_1 \hat{a}_1^\dagger \sigma_{\mathrm{fe}}^- + \frac{g_2}{\sqrt{2}} \hat{a}_2^\dagger \sigma_{\mathrm{eg}}^- + \frac{\Omega}{2\sqrt{2}} \sigma_{\mathrm{fe}}^- + H.c.\right),
\end{equation}
and the associated detunings defined by the rotated energy frame of $\hat{\mathcal{H}}_{\mathrm{A}} = A \sigma_{\mathrm{ee}} + A \sigma_{\mathrm{ff}} + A \hat{a}_1^\dagger \hat{a}_1$. Here, the extra terms which couple the ground and the excited states in the absence of a signal photon lead to non-zero dark count rate of the detector.

There are two separate dark count rates induced by the driving of the lower transition in the system as contained in Eq. (\ref{eq:Hsys}). The first involves a cycle $|g,0,0\rangle$-$|e,0,0\rangle$-$|g,0,1\rangle$-$|g,0,0\rangle$ with emission of a single photon at a rate $\Gamma_{\mathrm{dark}}^{\mathrm{(s)}}$. Importantly, this process does not include radiative transitions within the excited manifold. Thus it is not amplified by SPT, and can be discriminated for large SPT gains. The second process corresponds to ground-to-excited state transitions $|g,0,0\rangle$-$|e,0,0\rangle$-$|g,0,1\rangle$-$|e,0,1\rangle$-$|e,0,0\rangle$ happening at a rate $\Gamma_{\mathrm{dark}}^{\mathrm{(m)}}$, which projects the system to the excited manifold, launches a photon avalanche, and needs to be strongly suppressed.

To access the rates directly we find the full counting statistics using the wave function Monte-Carlo approach and study the no-jump evolution of the system [\onlinecite{SM}h], allowing us to estimate the dark count rates for the single and enhanced processes. The results are shown in Fig. \ref{fig:anharmonicity}(b). Additionally, the \emph{steady state} dark count rates can be calculated using an effective operator approach [\onlinecite{SM}h], valid for large anharmonicities, which gives
%
%\begin{tabularx}{\linewidth}{@{}XX@{}}
%  \begin{equation}
%  \Gamma_{\mathrm{dark,ss}}^{\mathrm{(s)}} \approx \frac{\kappa_2 g_2^2 \Omega^2}{4 (A^2 \kappa_2^2 + g_2^4)},\label{eq:Gamma_dark_S}
%  \end{equation} &
%  \begin{equation}
%  \Gamma_{\mathrm{dark,ss}}^{\mathrm{(m)}} \approx \frac{g_2^2 \Omega^4}{32 A^4 \kappa_2}.  \label{eq:Gamma_dark_M}
%  \end{equation}
%\end{tabularx}
%
%%%%%%
%
\begin{equation*}
\Gamma_{\mathrm{dark,ss}}^{\mathrm{(s)}} \approx \frac{\kappa_2 g_2^2 \Omega^2}{4 (A^2 \kappa_2^2 + g_2^4)},\quad (6)  \quad \Gamma_{\mathrm{dark,ss}}^{\mathrm{(m)}} \approx \frac{g_2^2 \Omega^4}{32 A^4 \kappa_2}. \quad (7)
\end{equation*}
While the simplified analytical result for $\Gamma_{\mathrm{dark,ss}}^{\mathrm{(s)}}$ coincides with numerical estimates for $A/g_2 > 40$, the enhanced dark count rate $\Gamma_{\mathrm{dark}}^{\mathrm{(m)}}$ shows an additional dynamical contribution due to induced jumps to the excited subspace triggered by the jumps with the rate $\Gamma_{\mathrm{dark,ss}}^{\mathrm{(s)}}$. From the numerical simulation we find that the total rate $\Gamma_{\mathrm{dark}}^{\mathrm{(m)}}$ retains the favorable $A^{-4}$ scaling but is roughly a factor of $\eta \approx 4$ larger than $\Gamma_{\mathrm{dark,ss}}^{\mathrm{(m)}}$ [\onlinecite{SM}h].

\textit{Real structure estimates.---}For a realistic example we consider a flux qubit where $g_2 = 2\pi\times 458$ MHz can be attained for an anharmonicity $A = 2\pi\times 8.426$ GHz and a decay rate $\gamma = 2\pi\times 0.227$ MHz \cite{Inomata2014}. Decreasing $g_2$ to $2\pi\times 120$ MHz, setting $\kappa_2 = g_2$, $g_1 = 2\pi \times 6$ MHz, and $\Omega = 2\pi \times 240$ MHz gives a gain of 172 photons with $2\pi\times 0.6$ MHz bandwidth, enhanced dark count rate of $2\pi \times 660$ Hz, and single dark count rate $2\pi \times 14.4$ kHz.

Finally, we note that for using the device as a single photon detector the gain photons needs to be measured. For the above scenario with a gain of approximately $200$ we estimate that the signal is distributed on $\sim 90$ modes. With a heterodyne detection setup this output field can be measured with an efficiency of $95\%$ with only a $0.02$ dark count probability and better performance can be achieved at higher gain [\onlinecite{SM}i]. Alternatively the expected signal begins to be within range of calorimetric detection schemes \cite{Govenius2016}.

\textit{Conclusion.---}We have presented a scheme for a single photon transistor based on the impedance-matched superconducting circuit, which operates in the cw regime and allows for an on-demand single microwave photon detection. The scheme can realistically lead to an output of several hundred photons, tolerates high pure dephasing rates, and keeps low dark count rates.

\textit{Acknowledgements.---}The research was funded by the European Union Seventh Framework Programme through the ERC Grant QIOS (Grant No. 306576).

\begingroup
\renewcommand{\addcontentsline}[3]{}% Remove functionality of \addcontentsline
\renewcommand{\section}[2]{}% Remove functionality of \section

\endgroup

\onecolumngrid
\newpage

\setcounter{equation}{0}
\setcounter{figure}{0}
\renewcommand{\theequation}{S\arabic{equation}}
\renewcommand{\thefigure}{S\arabic{figure}}

\begin{center}
\textbf{Supplemental Material: Continuous wave single photon transistor based on a superconducting circuit}
\end{center}

\tableofcontents

\subsection{\label{sect:A}Transformation to the rotating frame}

We start with the general Hamiltonian corresponding to a three level artificial atom coupled to two superconducting cavities, which are additionally linked to transmission line waveguides [see Fig. 1 in the main text for the geometry]. It can be divided into the system and waveguide terms. The original system Hamiltonian for large qubit anharmonicity reads: 
\begin{equation}
\label{eqS:Hsys_0}
\hat{\mathcal{H}}_{\mathrm{sys}} = \omega_{\mathrm{eg}} \sigma_{\mathrm{ee}} + (\omega_{\mathrm{eg}} +\omega_{\mathrm{fe}}) \sigma_{\mathrm{ff}} + \omega_{\mathrm{cav},1} \hat{a}_1^\dagger \hat{a}_1 + \omega_{\mathrm{cav},2} \hat{a}_2^\dagger \hat{a}_2 + g_1 (\hat{a}_1^\dagger \sigma_{\mathrm{eg}}^- + \sigma_{\mathrm{eg}}^+ \hat{a}_1) + g_2 (\hat{a}_2^\dagger \sigma_{\mathrm{fe}}^- + \sigma_{\mathrm{fe}}^+ \hat{a}_2) + \Omega (\sigma_{\mathrm{fe}}^- e^{i\omega_d t} + \sigma_{\mathrm{fe}}^+ e^{-i\omega_d t}),
\end{equation}
where $\omega_{\mathrm{eg}}$ corresponds to the $|e\rangle$-$|g\rangle$ energy difference of the qutrit ($\hbar = 1$), $\omega_{\mathrm{fe}} = \omega_{\mathrm{eg}} - A$ denotes the energy distance between $|f\rangle$ and $|e\rangle$ levels, defined by the anharmonicity of the qutrit levels, $A$. The energies of the cavities are  $\omega_{\mathrm{cav},1}$ and $\omega_{\mathrm{cav},2}$, correspondingly. $\omega_d$ refers to the frequency of the classical drive.
In Eq. (\ref{eqS:Hsys_0}) we have assumed the anharmonicity to be sufficiently large to suppress the classical drive between the $|e\rangle$-$|g\rangle$ states, as well as the cross-coupling between cavity 1 (2) and the upper (lower) artificial atomic transition. This restriction will be relaxed in  section \ref{sect:H} of this Supplemental Material.

The Hamiltonian for waveguides $j=1,2$ is
\begin{equation}
\label{eqS:Hwgd}
\hat{\mathcal{H}}_{\mathrm{wgd,j}} = \int_{-\infty}^{+\infty} dp \omega_{j,p} \hat{b}_{j,p}^\dagger \hat{b}_{j,p} - i \int_{-\infty}^{+\infty} dp \sqrt{\frac{\kappa_j}{2\pi}} \left( \hat{a}_j^\dagger \hat{b}_{j,p} - \hat{b}_{j,p}^\dagger \hat{a}_j \right),
\end{equation}
where $\hat{b}_{j,p}^\dagger$ ($\hat{b}_{j,p}$) denotes creation (annihilation) operator for waveguide mode $j=1,2$ with 1D wavevector $p$ ($c = 1$).

To transform the Hamiltonian to a suitable time-independent rotating frame we use a unitary transformation with the operator $\hat{U}=\exp(-i\hat{R}t)$, yielding $\hat{\mathcal{H}}' = \hat{U}^\dagger \hat{\mathcal{H}} \hat{U} - \hat{R}$. We choose the rotation operator as
\begin{equation}
\label{eqS:R}
\hat{R} = \omega_s |e\rangle \langle e| + \omega_s \hat{a}_1^\dagger \hat{a}_1 + (\omega_d + \omega_s) |f\rangle \langle f| + \omega_d \hat{a}_2^\dagger \hat{a}_2 + \omega_s \sum_k \hat{b}_{1,k}^\dagger \hat{b}_{1,k} + \omega_d \sum_k \hat{b}_{2,k}^\dagger \hat{b}_{2,k},
\end{equation}
where $\omega_s$ corresponds to the central frequency of the incident  single photon wavepacket. The reference frame is chosen to eliminate the time dependence of the  upper $|e\rangle$-$|f \rangle$ qutrit transition due to the drive, and to write the lower $|e\rangle$-$|g\rangle$ transition terms as a function of the input photon detuning.

The transformed system Hamiltonian thus reads [also in Eq. (2) of the main text]:
\begin{equation}
\label{eqS:Hsys}
\hat{\mathcal{H}}_{\mathrm{sys}} = \delta_e \sigma_{\mathrm{ee}} + (\delta_e + \delta_f) \sigma_{\mathrm{ff}} + \delta_{\mathrm{cav},1} \hat{a}_1^\dagger \hat{a}_1 + \delta_{\mathrm{cav},2} \hat{a}_2^\dagger \hat{a}_2 + g_1 (\hat{a}_1^\dagger \sigma_{\mathrm{eg}}^- + \sigma_{\mathrm{eg}}^+ \hat{a}_1) + g_2 (\hat{a}_2^\dagger \sigma_{\mathrm{fe}}^- + \sigma_{\mathrm{fe}}^+ \hat{a}_2) + \Omega (\sigma_{\mathrm{fe}}^- + \sigma_{\mathrm{fe}}^+),
\end{equation}
with detunings $\delta_e = \omega_{\mathrm{eg}} - \omega_s$, $\delta_f = \omega_{\mathrm{fe}} - \omega_d$, $\delta_{\mathrm{cav},1} = \omega_{\mathrm{cav},1} - \omega_s$, and $\delta_{\mathrm{cav},2} = \omega_{\mathrm{cav},2} - \omega_d$.

The transformed waveguide couplings read

\begin{align}
\label{eqS:Hwgd_12}
&\hat{\mathcal{H}}_{\mathrm{wgd,1}} = \int_{-\infty}^{+\infty} dp \delta_{1,p} \hat{b}_{1,p}^\dagger \hat{b}_{1,p} - i \int_{-\infty}^{+\infty} dp \sqrt{\frac{\kappa_1}{2\pi}} \left( \hat{a}_1^\dagger \hat{b}_{1,p} - \hat{b}_{1,p}^\dagger \hat{a}_1 \right),\\ \notag
&\hat{\mathcal{H}}_{\mathrm{wgd,2}} = \int_{-\infty}^{+\infty} dp \delta_{2,p} \hat{b}_{2,p}^\dagger \hat{b}_{2,p} - i \int_{-\infty}^{+\infty} dp \sqrt{\frac{\kappa_2}{2\pi}} \left( \hat{a}_2^\dagger \hat{b}_{2,p} - \hat{b}_{2,p}^\dagger \hat{a}_2 \right),
\end{align}
where $\delta_{1,p} = \omega_{1,p} - \omega_s$ and $\delta_{2,p} = \omega_{2,p} - \omega_d$. This choice of rotation frame redefines the coupling such that waveguide modes energies are relative to the input signal and drive frequencies.

%%%%%%%%%%%%%%%%%%%%%%%%%%%%%%%%%%%%%%%%%%%%%%%

\subsection{\label{sect:B}Diagonalization of the excited states manifold}

In order to describe the processes in the excited qutrit levels $|e\rangle$ and $|f\rangle$ it is convenient to truncate the full Hilbert space of the system to six levels $\{|g,0,0\rangle,|g,1,0\rangle,|e,0,0\rangle,|f,0,0\rangle,|e,0,1\rangle,|f,0,1\rangle\}$. In particular we are interested in the excited subspace $|\Psi_E\rangle = (|e,0,0\rangle,|f,0,0\rangle,|e,0,1\rangle,|f,0,1\rangle)^T$, with the states being efficiently mixed by cavity couplings and the classical drive. The system Hamiltonian projected onto the subspace of excited states can be written in matrix form as
\begin{equation}
\label{eqS:H_E}
\mathbf{H}_E =
\left( \begin{array}{cccc}
\delta_e & \Omega/2 & 0 & 0 \\
\Omega/2 & (\delta_e + \delta_f) & g_2 & 0 \\
0 & g_2 & (\delta_e + \delta_{\mathrm{cav},1}) & \Omega/2 \\
0 & 0 & \Omega/2 & (\delta_e + \delta_f + \delta_{\mathrm{cav},2}) \\
\end{array} \right).
\end{equation}
Considering the optimal zero detuning with $\delta_e = \delta_{\mathrm{cav},1} = 0$ and $\delta_f = \delta_{\mathrm{cav},2} = 0$, the matrix (\ref{eqS:H_E}) can be diagonalized in term of  dressed states $|\Psi_M \rangle = (|1\rangle,|2\rangle,|3\rangle,|4\rangle)^T$. The corresponding energies are:
\begin{align}
\label{eqS:E_dressed}
E_1 &= -\frac{g_2}{2} - \frac{\sqrt{g_2^2 + \Omega^2}}{2},\\ \notag
E_2 &= \frac{g_2}{2} - \frac{\sqrt{g_2^2 + \Omega^2}}{2},\\ \notag
E_3 &= -\frac{g_2}{2} + \frac{\sqrt{g_2^2 + \Omega^2}}{2},\\ \notag
E_4 &= \frac{g_2}{2} + \frac{\sqrt{g_2^2 + \Omega^2}}{2},
\end{align}
and the transformation between bare and dressed bases, $|\Psi_E\rangle = \mathbf{P} |\Psi_M \rangle$, can be performed using the matrix $\mathbf{P}$ which reads:
\begin{equation}
\label{eqS:P}
\mathbf{P} =
\left( \begin{array}{cccc}
-\beta & \alpha & -\alpha & \beta \\
\alpha & -\beta & -\beta & \alpha \\
-\alpha & -\beta & \beta & \alpha \\
\beta & \alpha & \alpha & \beta \\
\end{array} \right),
\end{equation}
where $\alpha$ and $\beta$ are constants defined as
\begin{equation}
\label{eqS:alpha_beta}
\alpha = \frac{1}{2}\sqrt{1 + \frac{g_2}{\sqrt{g_2^2 + \Omega^2}}}\quad \quad \mathrm{and}\quad \quad \beta = \frac{1}{2}\sqrt{1 - \frac{g_2}{\sqrt{g_2^2 + \Omega^2}}}.
\end{equation}
While coherent dynamics within the dressed excited subspace is trivial,  cavity 2 decay rewritten in terms of the new dressed states now leads to numerous jump terms between the levels. The original collapse operator for the second cavity $\hat{C}_{\kappa_2} = \sqrt{\kappa_2}\hat{a}_2 = \sqrt{\kappa_2} (|e,0,0\rangle \langle e,0,1| + |f,0,0\rangle \langle f,0,1|)$ can be recast as
\begin{align}
\label{eqS:C_kappa2}
\hat{C}_{\kappa_2,E} = &\sqrt{\kappa_2} \bigg[ \frac{\Omega}{2\sqrt{g_2^2 + \Omega^2}} \left(|1 \rangle\langle 1| - |2 \rangle\langle 2| - |3 \rangle\langle 3| + |4 \rangle\langle 4|\right) + \frac{1}{2} \left(|1 \rangle\langle 2| - |2 \rangle\langle 1| + |4 \rangle\langle 3| - |3 \rangle\langle 4|\right) \\ \notag &+ \frac{g_2}{2 \sqrt{g_2^2 + \Omega^2}} \left(|1 \rangle\langle 3| + |3 \rangle\langle 1| + |2 \rangle\langle 4| + |4 \rangle\langle 2|\right) \bigg],
\end{align}
and the possible jump processes are depicted in Fig. 1(c) of the main text. The form of Eq. (\ref{eqS:C_kappa2}) implies that a single $\kappa_2$ jump event projects the system into superposition of dressed states.

Finally, the no-jump evolution associated to the $\kappa_2$ jump operator in the excited subspace, which is defined by the  $-i \hat{C}_{\kappa_2,E}^\dagger \hat{C}_{\kappa_2,E}/2$ non-Hermitian term, can be rewritten in the dressed state picture as
\begin{equation}
\label{eqS:CdagC}
-\frac{i}{2} \hat{C}_{\kappa_2,E}^\dagger \hat{C}_{\kappa_2,E} = -\frac{i \kappa_2}{4} (|1 \rangle\langle 1| + |2 \rangle\langle 2| + |3 \rangle\langle 3| + |4 \rangle\langle 4|).
\end{equation}
We assumed the strong coupling regime, $\Omega, g_2 \gg \kappa_2$, which allows neglecting off-diagonal dissipative couplings. The corresponding non-Hermitian Hamiltonian then reads
\begin{equation}
\label{eqS:HNH}
\hat{\mathcal{H}}_{E}^{\mathrm{(NH)}} = \sum\limits_{j=1}^{4} (E_j - i \frac{\kappa_2}{4}) |j\rangle \langle j|.
\end{equation}
%

%%%%%%%%%%%%%%%%%%%%%%%%%%%%%%%%%%%%%%%%%%%%%%%

\subsection{\label{sect:C}Simple introduction to impedance matching}

As the important part of the proposed single photon transistor (and detector)  relies on the impedance matching (IM) concept, we shall provide a short introduction to IM using a few examples.
%%%
\begin{figure}
\includegraphics[width=0.75\linewidth]{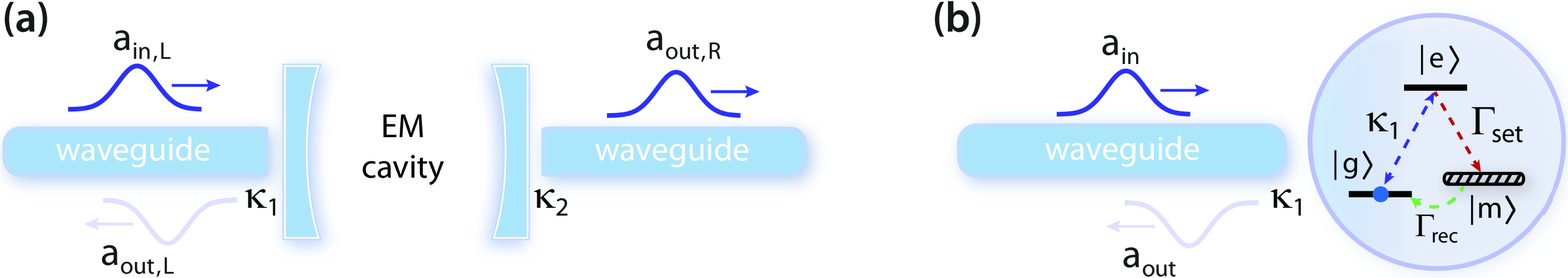}
\caption{(a) Double side coupled cavity with coupling to waveguides given by rates $\kappa_1$ and $\kappa_2$. For $\kappa_1 = \kappa_2$ an incident beam is always transmitted. (b) A three level $\Lambda$ atom coupled to a single waveguide through $|e\rangle$-$|g\rangle$ transition. Setting the decay rate $\Gamma_{\mathrm{set}}$ to a metastable $|m\rangle$ equal to the waveguide coupling $\kappa_1=\Gamma_{\mathrm{set}}$ allows impedance matching with full absorption of an incident pulse.}
\label{figS:IM}
\end{figure}
%%%

%%%%%
\subsubsection{Double-sided cavity}

The first and the simplest example where impedance matching appears in quantum optical context is a double-sided cavity [see sketch in Fig. \ref{figS:IM}(a)]. For this, the system Hamiltonian reads $\hat{\mathcal{H}}_{\mathrm{cav}} = \omega_{\mathrm{cav}} \hat{a}^\dagger \hat{a}$, and the waveguide Hamiltonian with couplings is
\begin{equation}
\hat{\mathcal{H}}_{\mathrm{cpl}} = \sum\limits_q \omega_q \hat{b}_{L,q}^\dagger \hat{b}_{L,q} + \sum\limits_q \omega_q \hat{b}_{R,q}^\dagger \hat{b}_{R,q} -i \sum\limits_q (f_{L,q} \hat{b}_{L,q}^\dagger \hat{a} - f_{L,q}^* \hat{a}^\dagger \hat{b}_{L,q}) -i \sum\limits_q (f_{R,q} \hat{b}_{R,q}^\dagger \hat{a} - f_{R,q}^* \hat{a}^\dagger \hat{b}_{R,q}),
\end{equation}
where $\hat{b}_{L,q}$ and $\hat{b}_{R,q}$ are bosonic waveguide modes and the coefficients $f_{L,q}$, $f_{R,q}$ denote mirror couplings. Following the standard input-output procedure (e.g. Ref. \cite{Clerk2010SM}, Supplemental Material, sec. E2; Refs. \cite{WallsMilburnSM,Fan2010SM}), the Heisenberg equation of motion (EOM) can be derived for the cavity annihilation operator $\hat{a}(t)$, yielding:
\begin{equation}
\label{eqS:a_HEOM_cav}
\dot{\hat{a}} = -i \omega_{\mathrm{cav}} \hat{a} - \frac{\kappa_1}{2} \hat{a} + \sqrt{\kappa_1} \hat{a}_{\mathrm{in},L}(t) - \frac{\kappa_2}{2} \hat{a} + \sqrt{\kappa_2} \hat{a}_{\mathrm{in},R}(t),
\end{equation}
where $\kappa_{1,2}$ now describe rates for left and right mirror couplings. The corresponding input-output relations read
\begin{equation}
\label{eqS:in-out_cav}
\hat{a}_{\mathrm{out},L}(t) = -\hat{a}_{\mathrm{in},L}(t) + \sqrt{\kappa_1} \hat{a}(t),\quad 
\hat{a}_{\mathrm{out},R}(t) = -\hat{a}_{\mathrm{in},R}(t) + \sqrt{\kappa_2} \hat{a}(t),
\end{equation}
with $\hat{a}_{\mathrm{in/out},L/R}$ being the input and output modes operators for left- and right-going waveguides. In the following we consider the input signal to enter from the left side and set $\langle \hat{a}_{\mathrm{in},R} \rangle = 0$. Eq. (\ref{eqS:a_HEOM_cav}) can be solved in the Fourier domain to give
\begin{equation}
\hat{a}(\omega) = \frac{\sqrt{\kappa_1} \hat{a}_{\mathrm{in},L}(\omega)}{(\kappa_1 + \kappa_2)/2 - i(\omega - \omega_{\mathrm{cav}})}.
\end{equation}
Considering a coherent cw input and using relation (\ref{eqS:in-out_cav}), the amplitude corresponding to the mean value of the output field $\langle \hat{a}_{\mathrm{out},L} \rangle$ can be found as
\begin{equation}
\langle \hat{a}_{\mathrm{out},L}(\omega) \rangle = \frac{\kappa_1 \alpha_{\mathrm{in},L}}{(\kappa_1+\kappa_2)/2 -i(\omega - \omega_{\mathrm{cav}})} - \alpha_{\mathrm{in},L},
\end{equation}
where $\langle \hat{a}_{\mathrm{in},L}(\omega) \rangle \equiv \alpha_{\mathrm{in},L}$ denotes the amplitude of the cw drive. Finally, the reflection coefficient for the left mirror at zero input signal detuning ($\omega = \omega_{\mathrm{cav}}$) can be written as
\begin{equation}
\label{eqS:refl_cav}
|r|^2 = \left| \frac{\kappa_1 - \kappa_2}{\kappa_1 + \kappa_2} \right|^2.
\end{equation}
This shows the absence of reflection for $\kappa_1 = \kappa_2$, signifying an impedance matched system.

%%%%%
\subsubsection{Three level $\Lambda$ atom}

Alternatively to the double-sided geometry, one can consider a three level emitter (atom or superconducting qutrit) coupled to a single waveguide [Fig. \ref{figS:IM}(b)]. This system was  shown to allow for full single photon absorption and long-distance entanglement of spin qubits \cite{Pinotsi2008SM}. The waveguide mode is coupled to the $|e\rangle$-$|g\rangle$ transition, and can decay to a metastable state $|m\rangle$ with a rate $\Gamma_{\mathrm{set}}$. The system Hamiltonian then reads:
\begin{equation}
\label{eqS:H_Lambda}
\hat{\mathcal{H}}_{\Lambda} = \omega_{e} \sigma_{\mathrm{ee}} + \omega_{m} \sigma_{\mathrm{mm}} + \sum\limits_q \omega_q \hat{b}_{q}^\dagger \hat{b}_{q} -i \sum\limits_q (f_{q} \hat{b}_{q}^\dagger \sigma_{\mathrm{eg}}^- - f_{q}^* \sigma_{\mathrm{eg}}^+ \hat{b}_{q}),
\end{equation}
where $\sigma_{\mathrm{em}}^+ = |e\rangle \langle m|$ ($\sigma_{\mathrm{em}}^- = |m\rangle \langle e|$), $\omega_e$ corresponds to the energy of the $|e\rangle$-$|g\rangle$ transition, and $\omega_{m}$ corresponds to the energy of the metastable level. The corresponding input-output EOMs can be derived similarly to the double-sided cavity case \cite{Fan2010SM}. Additionally, the atomic decay to the metastable state (the $|e\rangle \rightarrow |m\rangle$ process) can be introduced to the EOM of an arbitrary system operator $\hat{\mathcal{O}}$ in the form: $\dot{\hat{\mathcal{O}}}_{\mathrm{dec}} = \hat{C}^\dagger \hat{\mathcal{O}} \hat{C} - \{ \hat{C}^\dagger \hat{C}, \hat{\mathcal{O}} \}/2$, where the collapse operator for the process is $\hat{C} = \sqrt{\Gamma_{\mathrm{set}}} \sigma_{\mathrm{em}}^-$, with $\Gamma_{\mathrm{set}}$ being the decay rate. The corresponding system of equations reads:
\begin{align}
\label{eqS:EOM_Lambda_ee}
\dot{\sigma}_{\mathrm{ee}} &= \sqrt{\kappa_1} (\sigma_{\mathrm{eg}}^+ \hat{a}_{\mathrm{in}} + \hat{a}_{\mathrm{in}}^\dagger \sigma_{\mathrm{eg}}^-) - \kappa_1 \sigma_{\mathrm{ee}} - \Gamma_{\mathrm{set}} \sigma_{\mathrm{ee}},\\
\label{eqS:EOM_Lambda_eg}
\dot{\sigma}_{\mathrm{eg}}^- &= -i \omega_e \sigma_{\mathrm{eg}}^- -\sqrt{\kappa_1} (\sigma_{\mathrm{ee}} - \sigma_{\mathrm{gg}}) \hat{a}_{\mathrm{in}} - \frac{\kappa_1}{2} \sigma_{\mathrm{eg}}^- - \frac{\Gamma_{\mathrm{set}}}{2} \sigma_{\mathrm{eg}}^-,
\end{align}
where we have defined the atom-waveguide coupling rate $\kappa_1$ and the input mode $\hat{a}_{\mathrm{in}}$, with continuity relation
\begin{equation}
\hat{a}_{\mathrm{out}} = -\hat{a}_{\mathrm{in}} + \sqrt{\kappa_1}\sigma_{\mathrm{eg}}^- .
\end{equation}
Considering the weak excitation limit the nonlinear term can be simplified to $\langle (\sigma_{ee} - \sigma_{gg}) \hat{a}_{\mathrm{in}} \rangle \approx - \langle \hat{a}_{\mathrm{in}} \rangle$, effectively decoupling Eq. (\ref{eqS:EOM_Lambda_ee}) from the system. Then, for a coherent state input $\alpha_{\mathrm{in}}e^{-i\omega t}$ the outgoing amplitude yields
\begin{equation}
\label{eqS:a_out_Lambda}
\langle \hat{a}_{\mathrm{out}} \rangle = \frac{\kappa_1 \alpha_{\mathrm{in}}}{(\kappa_1 + \Gamma_{\mathrm{set}})/2 -i (\omega - \omega_e)} - \alpha_{\mathrm{in}},
\end{equation}
and the reflection coefficient (or equivalently transmission coefficient in a chiral geometry \cite{Witthaut2010SM}) at zero detuning can be written as
\begin{equation}
\label{eqS:r_Lambda}
|r|^2 = \left| \frac{\kappa_1 - \Gamma_{\mathrm{set}}}{\kappa_1 + \Gamma_{\mathrm{set}}} \right|^2,
\end{equation}
showing a vanishing outgoing signal for the  impedance matching condition $\kappa_1 = \Gamma_{\mathrm{set}}$. Eq. (\ref{eqS:r_Lambda}) implies that an incoming signal is fully transformed to an atomic excitation, and due to the $|e\rangle \rightarrow |m\rangle$ decay sets the system in the metastable state with unity probability.

Finally, we note that the similar conclusion holds if a single state $|m\rangle$ is substituted with a set of states $\mathcal{M}$, where probability to decay the atom being in state $|M_j\rangle$ is given by the branching ratio $\Gamma_{\mathrm{set},j}/\sum_j \Gamma_{\mathrm{set,j}}$.

%%%%%%%%%%%%%%%%%%%%%%%%%%%%%%%%%%%%%%%%%%%%%%%

\subsection{\label{sect:D}Derivation of the effective setting rate}

The setting rate is defined by the effective rate of going from $| g,1,0 \rangle$ state to any of the excited states $|\Psi_E\rangle$, followed by a quantum jump with the emission of a cavity 2 photon. The corresponding process can be described by an effective jump operator, derived with the adiabatic elimination procedure described in Ref. \cite{Reiter2012SM}. The associated collapse operator reads
\begin{equation}
\label{eqS:L_kappa2}
\hat{L}_{\kappa_2}^{\mathrm{eff}} = \hat{C}_{\kappa_2} \left[ \hat{\mathcal{H}}_{E}^{\mathrm{(NH)}} \right]^{-1} \hat{V}_E^+,
\end{equation}
where $\hat{\mathcal{H}}_{E}^{\mathrm{(NH)}}$ denotes non-Hermitian Hamiltonian for the excited state subspace and $\hat{V}_E^+$ is the excitation operator from the ground to the excited subspace. Finally, $\hat{C}_{\kappa_2}$ represents a cavity 2 photon jump within excited subspace.

In the following we prefer to use the bare state basis and do not restrict the Hilbert space to a single excitation, but truncate the cavity 2 Fock space at a level $\mathcal{N}_2$. The collapse operator acting in the excited subspace $\hat{C}_{\kappa_2}$ can then be expanded as 
\begin{equation}
\label{eqS:C_kappa_2_N2}
\hat{C}_{\kappa_2} = \sqrt{\kappa_2}\hat{a}_2 = \sqrt{\kappa_2} \sum\limits_{n_2 = 1}^{\mathcal{N}_2} \sqrt{n_2} (|e,0,n_2-1\rangle \langle e,0,n_2| + |f,0,n_2-1\rangle \langle f,0,n_2|),
\end{equation}
and the non-Hermitian Hamiltonian for the bare excited states at zero detuning reads:
\begin{align}
\label{eqS:HNH_E}
\hat{\mathcal{H}}_{E}^{\mathrm{(NH)}} = &\sum\limits_{n_2 = 0}^{\mathcal{N}_2} \left(0-i n_2 \frac{\kappa_2}{2}\right) \left( |e,0,n_2\rangle \langle e,0,n_2| + |f,0,n_2\rangle \langle f,0,n_2| \right) + \sum\limits_{n_2 = 0}^{\mathcal{N}_2} \frac{\Omega}{2} \left( |e,0,n_2\rangle \langle f,0,n_2| + |f,0,n_2\rangle \langle e,0,n_2| \right) \\ \notag
&+ \sum\limits_{n_2 = 1}^{\mathcal{N}_2} \sqrt{n_2} (|f,0,n_2-1\rangle \langle e,0,n_2| + |e,0,n_2\rangle \langle f,0,n_2-1|).
\end{align}
The excitation operator is responsible for the perturbative coupling of $|g,1,0\rangle$ to the excited subspace, and is given by $\hat{V}_E^+ = g_1 |e,0,0\rangle \langle g,1,0|$.
Finally, using (\ref{eqS:L_kappa2}), the setting rate can be calculated as the rate of emitting photons by the decay of cavity 2:
\begin{equation}
\Gamma_{\mathrm{set}} =  \langle g,1,0| \hat{L}_{\kappa_2}^{\mathrm{eff}\dagger} \hat{L}_{\kappa_2}^{\mathrm{eff}} | g,1,0 \rangle .
\end{equation}
The setting rate can be straightforwardly evaluated by symbolic inversion of the matrix (\ref{eqS:HNH_E}) and vector multiplication, fixing $\mathcal{N}_2$ to a certain value. The number of excitations at which the occupation of second cavity needs to be truncated is mainly defined by the drive frequency $\Omega$ and the decay rate of cavity 2, $\kappa_2$. For strong driving $\Omega/\kappa_2 \gg 1$ the dressed states become a superposition of states involving larger photons numbers $n_2$ and the procedure requires using a high value of $\mathcal{N}_2$. For strong decay  $\Omega/\kappa_2 < 1$ only low photon numbers $n_2$ are involved, and a lower value of $\mathcal{N}_2$ can be used.

First, let us derive results for a truncation at $\mathcal{N}_2 = 1$. This leads to a simple expression:
\begin{equation}
\label{eqS:Gset_1}
\Gamma_{\mathrm{set}}^{\mathcal{N}_2 = 1} = \frac{16 g_1^2 g_2^2 \kappa_2}{\kappa_2^2 \Omega^2 + \Omega^4},
\end{equation}
which is valid for small $g_1$ and $\Omega/\kappa_2 \ll 1$. Going to the two-photon manifold $\mathcal{N}_2 = 2$, the setting rate is modified to
\begin{equation}
\label{eqS:Gset_2}
\Gamma_{\mathrm{set}}^{\mathcal{N}_2 = 2} = \frac{16 g_1^2 g_2^2 \kappa_2 (16 g_2^2 + 4 \kappa_2^2 + \Omega^2)}{4 \kappa_2^2 \Omega^2 (4 g_2^2 + \kappa_2^2) + 5 \kappa_2^2 \Omega^4 + \Omega^6}.
\end{equation}
This expression was also given in the main text.
Finally, considering three cavity 2 excitations $\mathcal{N}_2 = 3$, one can derive the setting rate 
\begin{equation}
\label{eqS:Gset_3}
\Gamma_{\mathrm{set}}^{\mathcal{N}_2 = 3} = \frac{16 g_1^2 g_2^2}{\kappa_2} \left( \frac{1}{\Omega^2} - \frac{96 g_2^4 \kappa_2^2 \Omega^4}{f(g_2,\kappa_2,\Omega)^2} - \frac{72 g_2^2 \kappa_2^2 + 36 \kappa_2^4 + 13 \kappa_2^2 \Omega^2 + \Omega^4}{f(g_2,\kappa_2,\Omega)} \right),
\end{equation}
where we have used the definition
\begin{equation}
f(g_2,\kappa_2,\Omega) = 36 \kappa_2^2 (8 g_2^4 + 6 g_2^2 \kappa_2^2 + \kappa_2^4) + \kappa_2^2 \Omega^2 (88 g_2^2 + 49 \kappa_2^2) + 14 \kappa_2^2 \Omega^4 + \Omega^6.
\end{equation}

\begin{figure}
\includegraphics[width=0.6\linewidth]{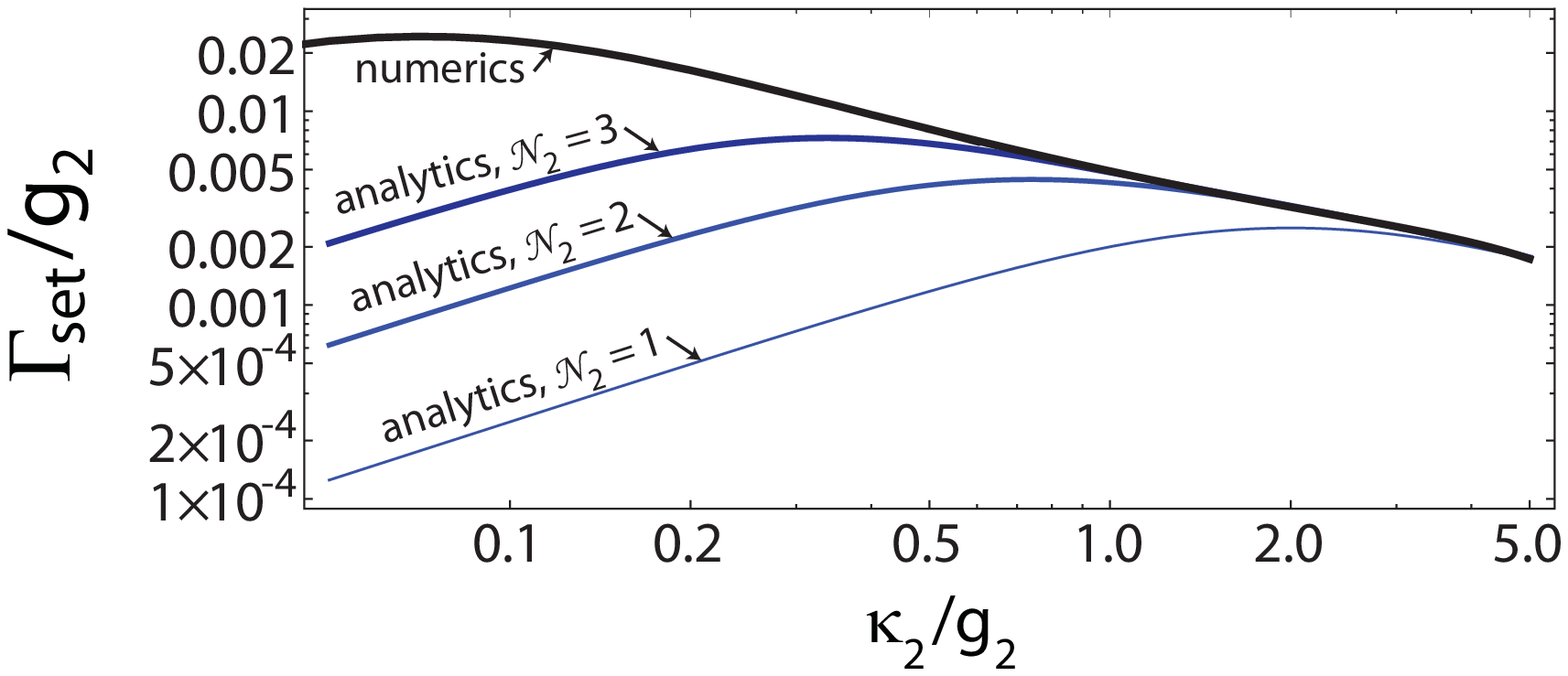}
\caption{Setting rate as a function of cavity 2 decay. The three blue curves correspond to analytical estimates which account for different maximal number of cavity 2 photons, $\mathcal{N}_2 = 1, 2, 3$. The black solid curve shows a full numerical result at high truncation $\mathcal{N}_2 = 10$, where results have converged. In the calculations we have used $\Omega/g_2 = 2$ and $g_1/g_2 = 0.05$.}
\label{figS:Gset}
\end{figure}

To find the applicability  for the analytical estimates of the setting rate given by Eqs. (\ref{eqS:Gset_1})-(\ref{eqS:Gset_3}), we plot in Fig. \ref{figS:Gset} its dependence on the decay rate of cavity 2, $\kappa_2/g_2$, together with a full numerical calculation, where we have used an $\mathcal{N}_2 = 10$ truncation.
%%%
%%%
We observe that the simple analytical formula (\ref{eqS:Gset_1}) obtained for a single cavity 2 excitation provides a correct estimate only for large cavity decay rate, $\kappa_2/g_2 \gtrsim 3$. However, already for a truncation $\mathcal{N}_2 = 2$ the analytical results give a good $\Gamma_{\mathrm{set}}$ estimate for $\kappa_2/g_2 \gtrsim 1.5$ range. Finally, the inclusion of three excitations allows to cover the behavior of  $\Gamma_{\mathrm{set}}$ down to a value $\kappa_2/g_2 \approx 0.6$.

With the   results in Eqs. (\ref{eqS:Gset_1})-(\ref{eqS:Gset_3}) we are thus able to give a good estimate of the setting rate, although with an increase in the complexity of the expression as the range of applicability grows. Importantly, all expressions predict a large degree of tunability with the drive strength $\Omega$. In an experiment this can therefore be used to fine tune the conditions for impedance matching, allowing the complete absorption of an incoming photon to the excited state manifold. 

%%%%%%%%%%%%%%%%%%%%%%%%%%%%%%%%%%%%%%%%%%%%%%%

\subsection{\label{sect:E}Heisenberg equations of motion for single photon input}

While most of the relevant information about the system can be obtained from the full master equation, the description of the incident single  photon  requires the exploitation of the input-output theory \cite{WallsMilburnSM}. This is conventionally done in the Heisenberg picture and relies on the derivation of Heisenberg equations of motion for system operators using the full system-bath Hamiltonian [Eq. (1), main text], and consequently a treatment of the bath modes as input or output of the actual system.

We start with the rotating frame Hamiltonian of the  system coupled to two input-output waveguides:
\begin{align}
\label{eqS:H_generic}
\hat{\mathcal{H}} = \hat{\mathcal{H}}_{\mathrm{sys}} + \sum\limits_{j=1,2} \Bigg[ \int_{-\infty}^{+\infty} dp \delta_{j,p} \hat{b}_{j,p}^\dagger \hat{b}_{j,p} - i \int_{-\infty}^{+\infty} dp \sqrt{\frac{\kappa_j}{2\pi}} \left( \hat{a}_j^\dagger \hat{b}_{j,p} - \hat{b}_{j,p}^\dagger \hat{a}_j \right) \Bigg],
\end{align}
where $\hat{\mathcal{H}}_{\mathrm{sys}}$ corresponds to the coupled qutrit-cavity Hamiltonian given by Eq. (2) of the main text. Here we assume the Markov approximation, implying that $\kappa_j$ is constant over the relevant range of frequencies. Using the input-output theory \cite{Clerk2010SM,WallsMilburnSM} we can account for external input from the waveguides, deriving the EOM for the system operators in the Heisenberg picture. Namely, the full Heisenberg equation for any system operator reads
\begin{equation}
\label{eqS:EOM_gen}
\dot{\hat{\mathcal{O}}} = -i [\hat{\mathcal{O}}, \hat{\mathcal{H}}_{\mathrm{sys}}] + \sum\limits_{j=1,2} \left( -\sqrt{\kappa_j} [\hat{\mathcal{O}},\hat{a}_j^\dagger] \hat{a}_{\mathrm{in},j} - \frac{\kappa_j}{2} [\hat{\mathcal{O}},\hat{a}_j^\dagger] \hat{a}_j + \sqrt{\kappa_j} \hat{a}_{\mathrm{in},j}^\dagger \left[ \hat{\mathcal{O}}, \hat{a}_j \right] + \frac{\kappa_j}{2} \hat{a}_j^\dagger \left[ \hat{\mathcal{O}}, \hat{a}_j \right] \right).
\end{equation}

Taking the explicit form of the system Hamiltonian (\ref{eqS:Hsys}), for example, the EOM for cavity 1 annihilation operator can be straightforwardly written as
\begin{equation}
\label{eqS:EOM_a1}
\dot{\hat{a}}_1 = -i \delta_{\mathrm{cav},1}\hat{a}_1 -i g_1 \sigma_{\mathrm{eg}}^- -\sqrt{\kappa_1} \hat{a}_{\mathrm{in},1} - \frac{\kappa_1}{2}\hat{a}_1.
\end{equation}

As we are interested only in the single input photon state, it is convenient to truncate the Fock space of the first cavity at the level of single excitation, $\mathcal{N}_1 = 1$. Then, the input process can be described with operators $|g,1,0\rangle \langle g,1,0| \equiv \hat{\sigma}_{11}$, $|g,0,0\rangle \langle g,0,0| \equiv \hat{\sigma}_{00}$, and $|g,0,0\rangle \langle g,1,0| \equiv \hat{\sigma}_{10}^{-}$. Using Eq. (\ref{eqS:EOM_gen}) the corresponding EOMs for the input sector can be written as
\begin{align}
\label{eqS:EOM_sigma11}
\dot{\hat{\sigma}}_{11} &= -\sqrt{\kappa_1} (\hat{\sigma}_{10}^{+} \hat{a}_{\mathrm{in},1} + \hat{a}_{\mathrm{in},1}^\dagger \hat{\sigma}_{10}^{-}) - \kappa_1 \hat{\sigma}_{11} -i [\hat{\sigma}_{11}, \hat{\mathcal{H}}_{\mathrm{sys}}],\\
\label{eqS:EOM_sigma10}
\dot{\hat{\sigma}}_{10}^{-} &= -\sqrt{\kappa_1} (\hat{\sigma}_{00}  - \hat{\sigma}_{11}) \hat{a}_{\mathrm{in},1} - \frac{\kappa_1}{2} \hat{\sigma}_{10}^- -i [\hat{\sigma}_{10}^-, \hat{\mathcal{H}}_{\mathrm{sys}}],
\end{align}
where last terms of Eqns. (\ref{eqS:EOM_sigma11})-(\ref{eqS:EOM_sigma10}) describe the coupling to the rest of the system.

To get the useful information about the system from the Heisenberg EOMs it is convenient to find matrix elements of the operators over the relevant input wave function \cite{Fan2010SM}. To this end we consider matrix elements  of the form $\langle \hat{\mathcal{O}} \rangle \equiv \langle\Psi_0 | \hat{\mathcal{O}} | \Psi_{\mathrm{in}} \rangle$, where $|\Psi_0 \rangle$ corresponds to the ground state of the atom, cavities and waveguides. The input state $| \Psi_{\mathrm{in}} \rangle$ is taken to be a momentum space eigenstate of the Hamiltonian written as
\begin{equation}
\label{eqS:Psi_in}
| \Psi_{\mathrm{in}} \rangle = \sum\limits_{m=g,e,f} \sum\limits_{n_1=0}^{\mathcal{N}_1} \sum\limits_{n_2=0}^{\mathcal{N}_2} C_{m,n_1,n_2} \frac{(\hat{a}_1^\dagger)^{n_1}}{\sqrt{n_1!}} \frac{(\hat{a}_2^\dagger)^{n_2}}{\sqrt{n_2!}} |m,{\O}\rangle + \int dk \alpha_{\mathrm{in},1}(k) \hat{b}_{1,k}^{\dagger}|g,{\O} \rangle,
\end{equation}
where $|m,{\O}\rangle = |m \rangle \otimes |{\O}\rangle$ with $|m\rangle$ being a qutrit state and $|{\O}\rangle = | 0_{\mathrm{cav},1}, 0_{\mathrm{cav},2}, 0_{\mathrm{wgd},1}, 0_{\mathrm{wgd},2} \rangle$ being the vacuum state for cavity modes and waveguide modes. $C_{m,n_1,n_2}$ are the amplitudes for the atom-cavities states. We consider the artificial atom to be in $|g\rangle$ prior to the single photon arrival. $\alpha_{\mathrm{in},1}(k)$ corresponds to the amplitude of the single photon incident on cavity 1, and we set the analogous term for the second input-output port to zero, $\alpha_{\mathrm{in},2}(k) \equiv 0$, implying the absence of input photons coming from the second waveguide.

We first consider the matrix element corresponding to input operator $\hat{a}_{1,\mathrm{in}}(t)$. Going to  momentum space, this operator can be rewritten as $\hat{a}_{\mathrm{in},1}(t) = \frac{1}{\sqrt{2\pi}} \int dk \hat{a}_{\mathrm{in},1}(k)e^{-i k t}$. Consequently, the relevant matrix element reads
\begin{equation}
\label{eqS:ain_av}
\langle g,{\O}|\hat{a}_{\mathrm{in},1}(t)|\Psi_{\mathrm{in}}\rangle = \langle g,{\O}| \frac{1}{\sqrt{2\pi}} \int dk \int dk' \alpha_{\mathrm{in},1}(k') e^{-i k t} \hat{a}_{\mathrm{in},1}(k) |g,1_{\mathrm{wgd},1,k'}\rangle = \langle g,{\O}| \frac{1}{\sqrt{2\pi}} \int dk \alpha_{\mathrm{in},1}(k) e^{-i k t} | g,{\O} \rangle.
\end{equation}
Assuming Gaussian single photon profile in the frequency domain,
\begin{equation}
\label{eqS:ain_k}
\alpha_{\mathrm{in},1}(k) = \frac{1}{\sqrt[4]{2\pi \sigma^2}} e^{-\frac{k^2}{4\sigma^2}},
\end{equation}
where $\sigma$ defines the frequency width of a pulse, the Fourier transform in Eq. (\ref{eqS:ain_av}) reduces to
\begin{equation}
\label{eqS:ain_t}
\alpha_{\mathrm{in},1}(t) = \langle g,{\O}|\hat{a}_{\mathrm{in},1}(t)|\Psi_{\mathrm{in}}\rangle =\sqrt[4]{\frac{2\sigma^2}{\pi}} e^{-\sigma^2 t^2}.
\end{equation}
We recall that in the rotating frame frequency $k$ is shifted by the carrier frequency of the pulse $k = k_0 - \omega_s$, thus removing fast oscillatoric time-dependence of the input pulse. It is also convenient to relate the frequency width $\sigma$ to the temporal width of the pulse $\tau = 1/(2\sigma)$.

Next, we write the EOM for the matrix elements describing the input sector. These are given by
\begin{align}
\label{eqS:EOM_rho11}
\langle \Psi_0| \dot{\hat{\sigma}}_{11} | \Psi_{\mathrm{in}}\rangle = \dot{\rho}_{11} &= -\sqrt{\kappa_1} [\rho_{01} \alpha_{\mathrm{in},1}(t) + \rho_{10} \alpha_{\mathrm{in},1}^*(t) ] - \kappa_1 \rho_{11} + \mathrm{coupling~terms},\\
\label{eqS:EOM_rho10}
\langle \Psi_0| \dot{\hat{\sigma}}_{10}^{-} | \Psi_{\mathrm{in}}\rangle = \dot{\rho}_{10} &= \sqrt{\kappa_1} \alpha_{\mathrm{in},1}(t) - \frac{\kappa_1}{2} \rho_{10} + \mathrm{coupling~terms},
\end{align}
where we used the property of single photon input, $\langle \Psi_0 |(\hat{\sigma}_{00}  - \hat{\sigma}_{11}) \hat{a}_{\mathrm{in},1}|\Psi_{\mathrm{in}}\rangle = -\langle \Psi_0 |\hat{a}_{1,\mathrm{in}}|\Psi_{\mathrm{in}}\rangle = -\alpha_{1,\mathrm{in}}(t)$. Here $\rho_{ij}$ denote elements of the density matrix. Finally, a closed system of $c$-number equations can be derived for the rest of system operators, leading to master-like equations, supplemented with normalization condition.

Using the density matrix equation for the system with incorporated single photon input we can calculate the occupation of cavity 2, and correspondingly the associated photon flux at the output of waveguide 2. This is used to plot Fig. 3(a) in the main text, and confirms perfect mapping of a flying input photon into a system excitation.

%%%%%%%%%%%%%%%%%%%%%%%%%%%%%%%%%%%%%%%%%%%%%%%

\subsection{\label{sect:F}Wave function Monte-Carlo calculations for single photon transistor statistics}

In the previous section we have derived Heisenberg equations of motion which allow to access averaged observables of the SPT system. Next, in order to better understand  the operation of the device and determine the counting statistics, it is instructive to use wave function Monte-Carlo (wfMC) calculations \cite{Dalibard1992SM,Molmer1993SM}. The method relies on the stochastic simulation of the system's wavefunction, accounting for possible quantum jumps inserted according to decay probabilities.
%%%
\begin{figure}
\includegraphics[width=0.5\linewidth]{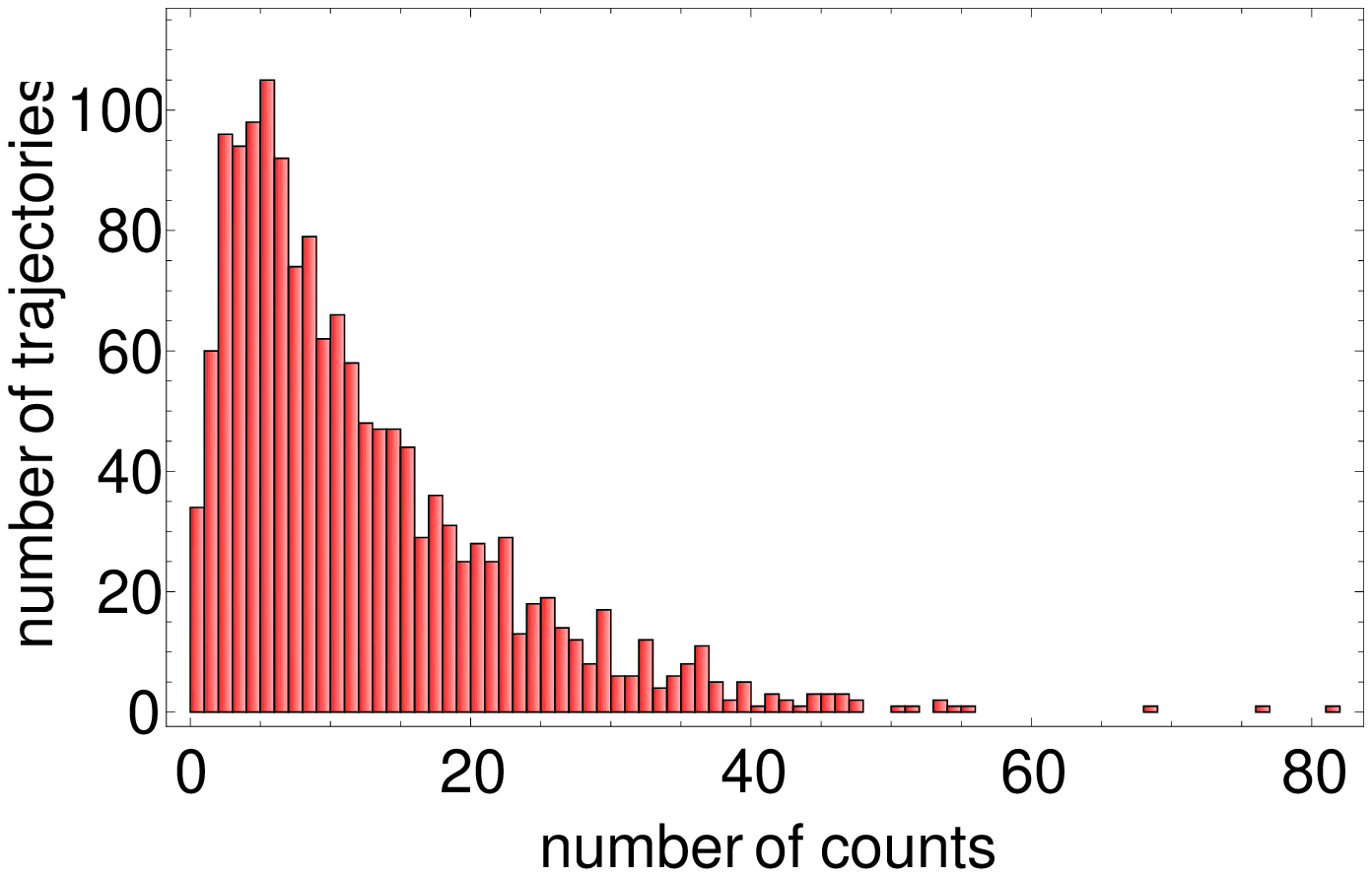}
\caption{Count distribution for output gain photons, plotted for $N_{\mathrm{ntraj}} = 1500$ trajectories. The parameters of the system are $\Omega/g_2 = 2$, $g_1/g_2 = 0.25$, $\kappa_2/g_2 = 1$, and $\kappa_1 = \Gamma_{\mathrm{set}}$. The average gain for the given parameters is approximately 12 photons.}
\label{figS:Pn}
\end{figure}
%%%

The general procedure for the simulation of the system dynamics is performed in the following way: First, the system Hamiltonian $\hat{\mathcal{H}}_{\mathrm{sys}}$ is truncated at fixed excitation levels $\mathcal{N}_1$ and $\mathcal{N}_2$, and the creation/annihilation operators are projected onto the  corresponding subspaces. The collapse operators corresponding to quantum jumps with emission of photons to the first and second waveguide are given by
\begin{align}
\label{eqS:C_kappa_1_full}
&\hat{C}_{\kappa_1} = \sqrt{\kappa_1}\hat{a}_1 = \sqrt{\kappa_1} \sum\limits_{m=g,e,f} \sum\limits_{n_1=1}^{\mathcal{N}_1} \sum\limits_{n_2 = 0}^{\mathcal{N}_2} \sqrt{n_1} |m,n_1-1,n_2\rangle \langle m,n_1,n_2|,\\
\label{eqS:C_kappa_2_full}
&\hat{C}_{\kappa_2} = \sqrt{\kappa_2}\hat{a}_2 = \sqrt{\kappa_2} \sum\limits_{m=g,e,f} \sum\limits_{n_1=0}^{\mathcal{N}_1} \sum\limits_{n_2 = 1}^{\mathcal{N}_2} \sqrt{n_2} |m,n_1,n_2-1\rangle \langle m,n_1,n_2|.
\end{align}

The wave function of the system in the Sch{\"o}dinger picture reads
\begin{equation}
\label{eqS:Psi_MC_0}
| \Psi(t) \rangle = \sum\limits_{m=g,e,f} \sum\limits_{n_1=0}^{\mathcal{N}_1} \sum\limits_{n_2=0}^{\mathcal{N}_2} c_{m,n_1,n_2}(t) \frac{(\hat{a}_1^\dagger)^{n_1}}{\sqrt{n_1!}} \frac{(\hat{a}_2^\dagger)^{n_2}}{\sqrt{n_2!}} |m,0,0\rangle + | \psi_{\mathrm{in}}(t)\rangle,
\end{equation}
with time-dependent state amplitudes $c_{m,n_1,n_2}(t)$ and single photon input part
\begin{equation}
\label{eqS:psi_in_t}
| \psi_{\mathrm{in}}(t)\rangle = \int dk \alpha_{1,\mathrm{in}}(t) \hat{b}_{1,k}^{\dagger}|g,{\O} \rangle.
\end{equation}

Next, the system is evolved under the non-Hermitian Hamiltonian given by
\begin{equation}
\hat{\mathcal{H}}_{\mathrm{NH}} = \hat{\mathcal{H}}_{\mathrm{sys}} - \frac{i}{2} \sum\limits_{j} \hat{C}_j^\dagger \hat{C}_j,
\end{equation}
where the Sch{\"o}dinger equation for the amplitudes $c_{m,n_1,n_2}(t)$ is used together with an input-output treatment of the single photon input [see Ref. \cite{Manzoni2014SM} and Supplemental Material therein for a discussion of the no-jump evolution]. The jump probabilities for different collapse operators $\hat{C}_j$ ($j = \kappa_1,\kappa_2$) per infinitesimal time span $\delta t$ are given by $\delta p_j / \delta t \equiv P_j = \langle \Psi(t)| \hat{C}_j^\dagger \hat{C}_j | \Psi(t)\rangle$. The jumps are inserted according to the normalized probabilities $P_j/\sum_j P_j$. The state after the jump collapses to $\hat{C}_j|\Psi(t)\rangle$.

Finally, an issue of high importance is the renormalization of the wave function, required to preserve the norm during the non-Hermitian evolution. As it involves the probability to be in the system's ground state while having an incoming photon in the waveguide, the total wave function norm reads
\begin{equation}
\label{eqS:Psi_norm}
\langle \Psi(t)|\Psi(t) \rangle = \sum\limits_{m=g,e,f} \sum\limits_{n_1=0}^{\mathcal{N}_1} \sum\limits_{n_2=0}^{\mathcal{N}_2} |c_{m,n_1,n_2}(t)|^2 + \int\limits_{t}^{+\infty} dt' I_{\mathrm{in},1}(t'),
\end{equation}
where we have defined the input intensity $I_{\mathrm{in},1}(t') = |\alpha_{\mathrm{in},1}(t')|^2$. Eq. (\ref{eqS:Psi_norm}) implies that for $t\rightarrow -\infty$ the wavefunction is normalized to be in the input state given by second term, while after the interaction it decreases to zero.

Solving the dynamical equations for $c_{m,n_1,n_2}(t)$ with the Monte-Carlo procedure for a large number of trajectories $N_{\mathrm{traj}}$ allows accessing the averages of any system operator $\hat{\mathcal{O}}$, with the results converging to the findings with the Heisenberg EOM  described before. Additionally, the stochastic wave function simulation strongly resembles real measurement, and allows to extract statistical properties of the outgoing photon flux. In particular, running $N_{\mathrm{traj}} = 1500$ calculations we plot the count distribution of output waveguide 2 photons in Fig. \ref{figS:Pn}. Each trajectory corresponds to a single cycle of the SPT operation triggered by single photon, which projects the system to the excited states manifold. Here, we deliberately increased the $g_1$ coupling to decrease the gain, while keeping large bandwidth $\kappa_1 = 0.12 g_2$. The Hilbert space is truncated at $\mathcal{N}_1 = 2$ and $\mathcal{N}_2 = 16$. The average number of output photons, being the gain of single photon transistor, is equal to $N_{\mathrm{out},2} = 11.67$, with the variance of the distribution $\Delta N_{\mathrm{out},2}^2 = 101$ largely exceeding the mean. Thus, the source exhibits the super-Poissonian statistics. The second-order coherence function at zero delay for the distribution can be calculated to be $g^{(2)}(0) = 1 + \frac{\Delta N_{\mathrm{out},2}^2 + N_{\mathrm{out},2}}{(N_{\mathrm{out},2})^2} = 1.66$ \cite{Davidovich1996SM}. This corresponds to the mixed statistics, which is in between coherent and thermal source behavior.
%%%
\begin{figure}%[h!]
\includegraphics[width=0.8\linewidth]{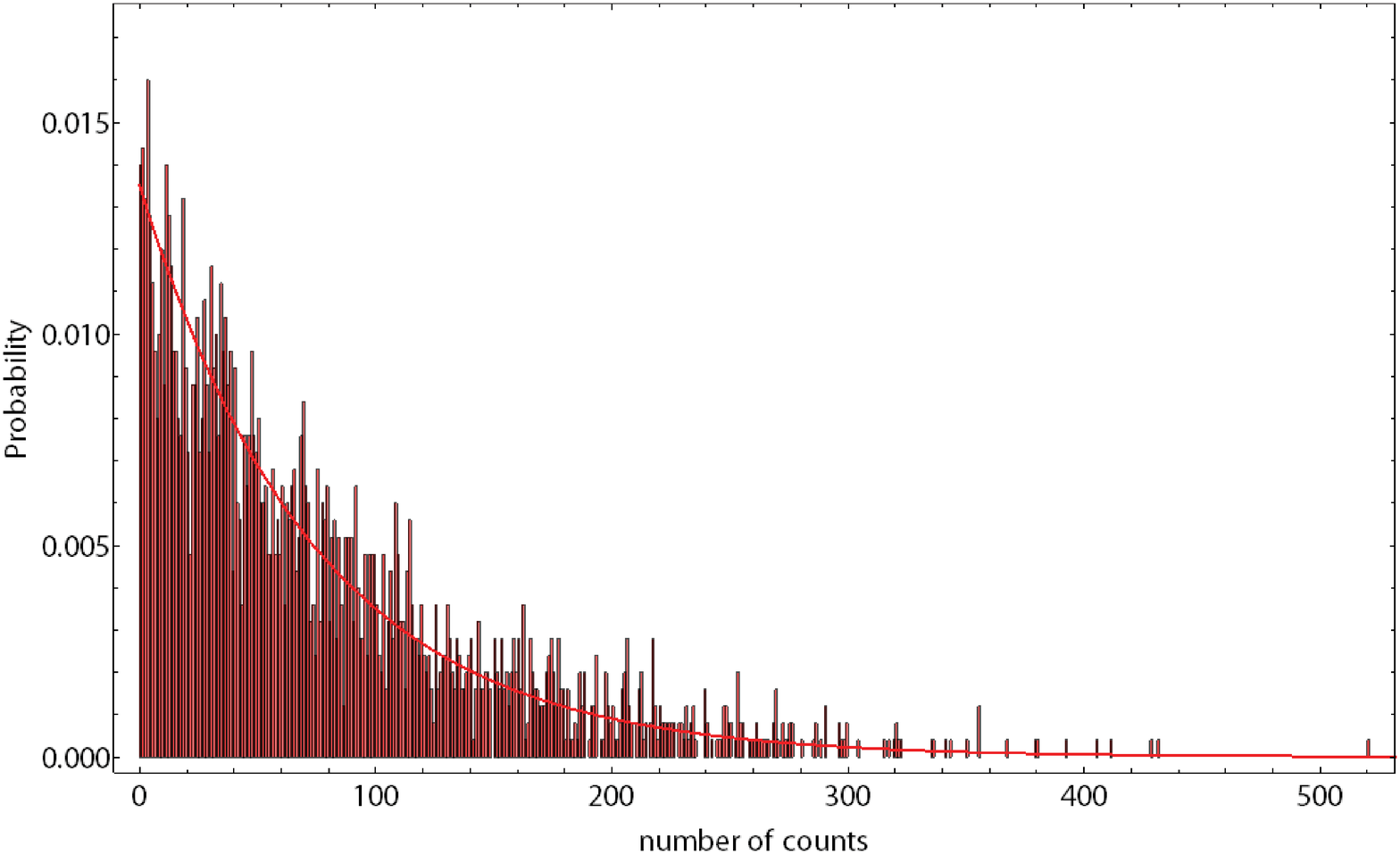}
\caption{Probability distribution for output gain photons, plotted for $N_{\mathrm{ntraj}} = 2500$ trajectories. The parameters of the system are $\Omega/g_2 = 1/2$, $\kappa_2/g_2 = 1$, $\kappa_1 = \Gamma_{\mathrm{set}}$, $g_1/g_2 = 0.005$, corresponding to a gain of $N_{\mathrm{out},2} \approx 74$. The solid curve corresponds to the exponential distribution $P(n)=\exp(-n/\lambda)/\lambda$ with $\lambda = N_{\mathrm{out},2}$.}
\label{figS:prob_large}
\end{figure}
%%%

As a probability distribution of outgoing photons is an important issue for the operation of SPT in the detector mode, we study its behavior in more details. In particular, we try to access the parameter range corresponding to large transistor gain. In general it is a formidable task. The calculation of the probability distribution requires a huge number of trajectories to access full statistical properties. Additionally, it involves numerical integration up to very long time, which increases dramatically for large gain. Assuming a moderate gain, we use the wave function Monte-Carlo approach with an initial state being a single photon in the excited subspace. Considering parameters $g_1/g_2 = 0.005$, $\Omega/g_2 = 1/2$, $\kappa_2/g_2 = 1$, which lead to a gain of $N_{\mathrm{out},2} \approx 73$ photons, we calculate the number of gain photons emitted within a time $t_{\mathrm{traj}} = 9000 g_2^{-1}$. This was found to be sufficiently long to collect $99.7\%$ of gain photons on average. The truncation of Hilbert space corresponds to $\mathcal{N}_1 = 4$ and $\mathcal{N}_2 = 16$. The calculated distribution of counts is shown in Fig. \ref{figS:prob_large}. The analysis reveals an approximately exponential behavior of the distribution, with a modification at the small count region ($<5$). We note that full access to the statistical properties is limited given the number of trajectories we simulated. The exponential distribution, however, corresponds to having photons leaving at a constant rate when the system is in the excited subspace combined with a fixed decay rate out of the subspace. It is thus natural to expect this exponential distribution apart from the very beginning, where the system needs a few photon emissions before it reaches a constant emission rate in the excited subspace.

%%%%%%%%%%%%%%%%%%%%%%%%%%%%%%%%%%%%%%%%%%%%%%%

\subsection{\label{sect:G}Definitions for decay and decoherence processes used in the imperfections analysis}

%%%%%
\subsubsection{Introduction of artificial atom decay and decoherence terms}

In the calculations of the SPT setting and output stage shown in Figs. 2 and 3 of the main text, we considered the qutrit decay to be small compared to the relevant coupling and cavity decay parameters. In Fig. 4(a) we test this assumption by adding both decay and decoherence to the $|e\rangle$-$|g\rangle$ and $|f\rangle$-$|e\rangle$ transitions. This can be done using the collapse operators
\begin{align}
\label{eqS:collapse_gamma_eg}
&\hat{C}_{\gamma_{\mathrm{eg}}} = \sqrt{\gamma_{\mathrm{eg}}} |g\rangle \langle e|,\\
\label{eqS:collapse_gamma_fe}
&\hat{C}_{\gamma_{\mathrm{fe}}} = \sqrt{\gamma_{\mathrm{fe}}} |e\rangle \langle f|,\\
\label{eqS:collapse_gamma_p_ee}
&\hat{C}_{\gamma_{\mathrm{p,ee}}} = \sqrt{\gamma_{\mathrm{p,ee}}} |e\rangle \langle e|,\\\label{eqS:collapse_gamma_p_ff}
&\hat{C}_{\gamma_{\mathrm{p,ff}}} = \sqrt{\gamma_{\mathrm{p,ff}}} |f\rangle \langle f|,
\end{align}
where $\gamma_{\mathrm{eg}}$ and $\gamma_{\mathrm{fe}}$ denote artificial atom decay rates for the lower and upper transitions, and $\gamma_{\mathrm{p,ee}}$, $\gamma_{\mathrm{p,ff}}$ correspond to the pure dephasing rates for the second and third qutrit levels. The collapse operators can be easily implemented into wave function Monte-Carlo calculations by adding extra terms describing  $|j\rangle \rightarrow |i\rangle$ ($j,i = f,e,g$) jumps. This causes additional norm decay by $-\frac{i}{2}\sum_j \hat{C}_{j}^\dagger \hat{C}_{j}$, with the index $j$ spanning all possible collapses (\ref{eqS:collapse_gamma_eg})-(\ref{eqS:collapse_gamma_p_ff}), with the corresponding probability to jump in a time interval $\delta t$ given by $\langle \psi(t)|\hat{C}_{j}^\dagger \hat{C}_{j}|\psi(t)\rangle \delta t$.

For the Heisenberg equations of motion the atomic decay and dephasing terms can be introduced through couplings to additional reservoirs modes $\hat{c}_{j,k}$ and $\hat{d}_{j,k}$. The decay of the artificial atom is then described by the Hamiltonian
\begin{equation}
\label{eq:Hwgd_gamma}
\hat{\mathcal{H}}_{j,\mathrm{decay}} = \int_{-\infty}^{+\infty} dk \omega_{j,k} \hat{c}_{j,k}^\dagger \hat{c}_{j,k} - i \int_{-\infty}^{+\infty} dk \sqrt{\frac{\gamma_j}{2\pi}}  ( \sigma_{j}^+ \hat{c}_{j,k} - \hat{c}_{j,k}^\dagger \sigma_{j}^- ),
\end{equation}
where the index $j=\mathrm{eg}$ ($\mathrm{fe}$) corresponds to the lower  (upper) transition.

The pure dephasing term emerges from shifts of the energy levels  of the qutrit, and is  given by
\begin{equation}
\label{eq:Hwgd_gamma_p}
\hat{\mathcal{H}}_{j',\mathrm{dephasing}} = \int_{-\infty}^{+\infty} dk \omega_{j',k} \hat{d}_{j',k}^\dagger \hat{d}_{j',k} - i \int_{-\infty}^{+\infty} dk \sqrt{\frac{\gamma_{\mathrm{p},j'}}{2\pi}}  (\hat{d}_{j',k} - \hat{d}_{j',k}^\dagger ) \sigma_{j'},
\end{equation}
where the index $j'=\mathrm{ee},\mathrm{ff}$ spans the second and third artificial atom levels. 

Finally, in the density matrix calculations atomic decay and dephasing are introduced by the Lindblad operator $\hat{\mathcal{D}}[\rho] = \hat{C}_{j} \rho \hat{C}_{j}^\dagger - \{ \hat{C}_{j}^\dagger \hat{C}_{j}, \rho \}/2$. With these results we thus account for the effect of decay and dephasing of the qutrit. In the main text this was used to calculate the reduction of SPT gain due to qutrit decay [Fig. 4(a)]. There the density matrix approach with collapse operators (\ref{eqS:collapse_gamma_eg})-(\ref{eqS:collapse_gamma_p_ff}) was used. Eqs. (\ref{eq:Hwgd_gamma})-(\ref{eq:Hwgd_gamma_p}) were used to include the decay and pure dephasing of the qubit into Heisenberg equations, allowing to check the impedance matching conditions with imperfections.

%%%%%
\subsubsection{Rate definitions}

To separate the influence of atomic decay channel and pure decoherence effects we consider them individually in  Fig. 4(a) of the main text. The solid curves are shown for negligibly dephasing rates $\gamma_{\mathrm{p,ee}}=\gamma_{\mathrm{p,ff}}=0$. Here we assume the ratio of qutrit decay rates to be fixed and defined using single parameter $\gamma$:
\begin{equation}
\gamma_{\mathrm{eg}} \equiv \gamma \quad \quad \mathrm{and} \quad \quad \gamma_{\mathrm{fe}} \equiv 2\gamma.
\end{equation}
This corresponds to the situation where the system is close to being a harmonic oscillator such that the matrix element for the upper transition is larger by a factor of $\sqrt{2}$.  
The dashed curves in Fig. 4(a) depict the situation, where the qutrit decay rates are relatively small (i.e. $\gamma_{\mathrm{eg}},\gamma_{\mathrm{fe}} \ll 10^{-4} g_2$), while the pure dephasing rates are substantial. This type of decoherence is especially relevant for superconducting circuits with large anharmonicity (small charging to Josephson energy ratio) \cite{Koch2007SM}. For the calculations we set $\gamma_{\mathrm{eg}}/g_2 \rightarrow 0,\gamma_{\mathrm{fe}}/g_2 \rightarrow 0$, define 
\begin{equation}
\gamma_{\mathrm{p,eg}} \equiv \gamma \quad \quad \mathrm{and} \quad \quad \gamma_{\mathrm{p,fe}} \equiv 2\gamma,
\end{equation}
and use $\gamma/g_2$ as a dimensionless variable for the plot.

%%%%%%%%%%%%%%%%%%%%%%%%%%%%%%%%%%%%%%%%%%%%%%%

\subsection{\label{sect:H}Dark count rate}

In the previous sections we considered an infinitely large anharmonicity $A$ for the superconducting artificial atom, which allowed us to retain only resonant couplings on the lower and upper qutrit transitions. However, this approximation breaks down if the value of anharmonicity becomes comparable to the coupling parameters. The full Hamiltonian of the system accounting for the finite $A$ can be written as
\begin{align}
\label{eqS:Hsys_0_A}
\hat{\mathcal{H}}_{\mathrm{sys},A} = &0\cdot \sigma_{\mathrm{gg}} + \omega_{\mathrm{eg}} \sigma_{\mathrm{ee}} + (2\omega_{\mathrm{e}} - A) \sigma_{\mathrm{ff}} + \omega_{\mathrm{cav},1} \hat{a}_1^\dagger \hat{a}_1 + \omega_{\mathrm{cav},2} \hat{a}_2^\dagger \hat{a}_2 + g_1 (\hat{a}_1^\dagger \sigma_{\mathrm{eg}}^- + \sigma_{\mathrm{eg}}^+ \hat{a}_1) + g_2 (\hat{a}_2^\dagger \sigma_{\mathrm{fe}}^- + \sigma_{\mathrm{fe}}^+ \hat{a}_2)  \\ \notag
 &+ \Omega (\sigma_{\mathrm{fe}}^- e^{i\omega_d t} + \sigma_{\mathrm{fe}}^+ e^{-i\omega_d t}) + \sqrt{2} g_1 (\hat{a}_1^\dagger \sigma_{\mathrm{fe}}^- + \sigma_{\mathrm{fe}}^+ \hat{a}_1) + \frac{g_2}{\sqrt{2}} \left(\hat{a}_2^\dagger \sigma_{\mathrm{eg}}^- + \sigma_{\mathrm{eg}}^+ \hat{a}_2 \right) + \frac{\Omega}{\sqrt{2}} \left( \sigma_{\mathrm{eg}}^- e^{i\omega_d t} + \sigma_{\mathrm{eg}}^+ e^{-i\omega_d t} \right),
\end{align}
where we have used that the anharmonicity typically reduces the frequency of the upper $|e \rangle$-$|f \rangle$ transition such that its transition frequency is $\omega_{\mathrm{fe}} = \omega_{\mathrm{eg}} - A$. The last three terms correspond to residual couplings arising from the fact that the three-level artificial atom is represented by an anharmonic oscillator. Since the anharmonic oscillator is typically not very far from being harmonic, we have for simplicity fixed the ratio of the transition matrix elements between the lower and upper transitions to be $1:\sqrt{2}$, and we set the values of the resonant couplings $g_1$ and $g_2$ as a reference.

An important difference between the Hamiltonian (\ref{eqS:Hsys_0_A}) and the one for an infinite $A$  (\ref{eqS:Hsys_0}) is given by the residual classical drive of the lower transition $|g \rangle$-$|e \rangle$. This allows for the excitation of the system even in the absence of incoming signal photon, and thus contributes to the dark count rate of the detector. To account for this process, we consider the situation without an incoming single photon and transform the system Hamiltonian to a suitable frame given by rotation operator
\begin{equation}
\label{eqS:R_A}
\hat{R} = \omega_d |e\rangle \langle e| + \omega_d \hat{a}_1^\dagger \hat{a}_1 + 2 \omega_d |f\rangle \langle f| + \omega_d \hat{a}_2^\dagger \hat{a}_2.
\end{equation}
This transformation yields the time-independent Hamiltonian
\begin{align}
\label{eqS:Hsys_A}
\hat{\mathcal{H}}_{\mathrm{sys},A} = &0\cdot \sigma_{\mathrm{gg}} + (\Delta + A) \sigma_{\mathrm{ee}} + (2 \Delta + A) \sigma_{\mathrm{ff}} + (\Delta + A + \delta_1) \hat{a}_1^\dagger \hat{a}_1 + (\Delta + \delta_2) \hat{a}_2^\dagger \hat{a}_2 + g_1 (\hat{a}_1^\dagger \sigma_{\mathrm{eg}}^- + \sigma_{\mathrm{eg}}^+ \hat{a}_1) \\ \notag
 &+ g_2 (\hat{a}_2^\dagger \sigma_{\mathrm{fe}}^- + \sigma_{\mathrm{fe}}^+ \hat{a}_2) + \Omega (\sigma_{\mathrm{fe}}^- + \sigma_{\mathrm{fe}}^+ ) + \sqrt{2} g_1 (\hat{a}_1^\dagger \sigma_{\mathrm{fe}}^- + \sigma_{\mathrm{fe}}^+ \hat{a}_1) + \frac{g_2}{\sqrt{2}} \left(\hat{a}_2^\dagger \sigma_{\mathrm{eg}}^- + \sigma_{\mathrm{eg}}^+ \hat{a}_2 \right) + \frac{\Omega}{\sqrt{2}} \left( \sigma_{\mathrm{eg}}^- + \sigma_{\mathrm{eg}}^+ \right),
\end{align}
where we have defined the small detunings $\Delta = \omega_{\mathrm{fe}} - \omega_d = \omega_{\mathrm{eg}} - A - \omega_d$, $\delta_1 = \omega_{\mathrm{cav},1} - \omega_{\mathrm{eg}}$, $\delta_2 = \omega_{\mathrm{cav,2}} - \omega_{\mathrm{fe}} = \omega_{\mathrm{cav},2} - \omega_{\mathrm{eg}} + A$. In the following we consider all detunings to be zero, $\Delta = \delta_1 = \delta_2 = 0$, corresponding to resonant driving.
%%%
\begin{figure}
\includegraphics[width=0.75\linewidth]{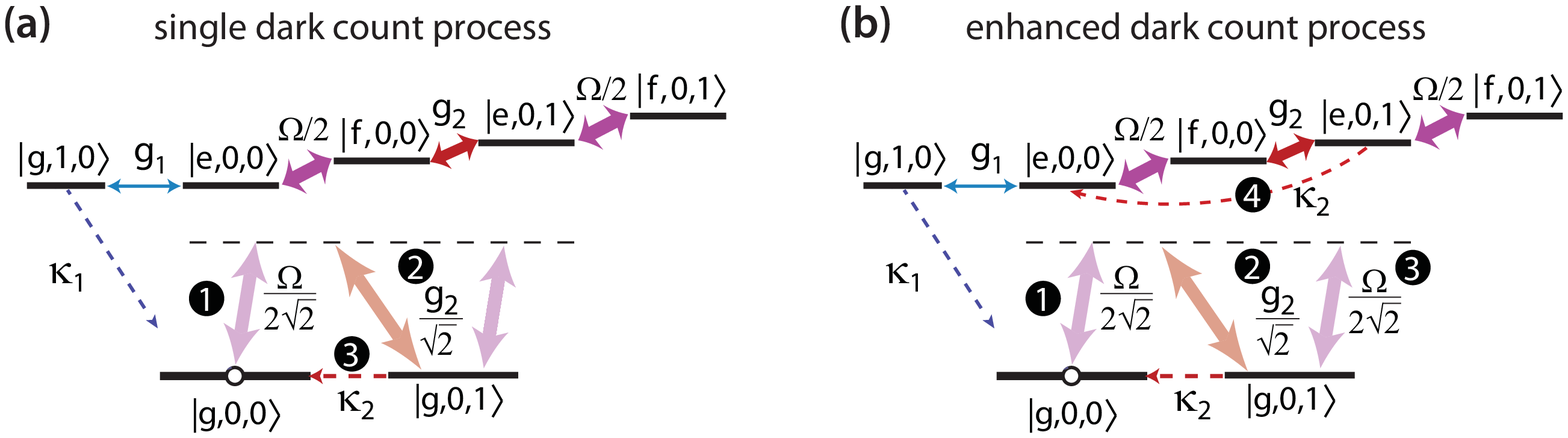}
\caption{Dark count processes of the SPT. (a) A single dark count cycle triggered by a radiative cavity 2 transition which returns the system to the ground state. The relevant sequence leading to steady state single dark counts are depicted by consequent bullets (1), (2), and (3). Here, a violet arrow corresponds to a classical drive transition, a thick red arrow denotes cavity 2 associated transition, and a thin blue arrow shows the coupling to the first cavity. The red and black dashed arrows correspond to cavity 2 and cavity 1 decay processes, respectively. The dashed horizontal line represents an anharmonicity of the qutrit. (b) Enhanced dark count process which projects the system to the excited subspace and mimics the arrival of a signal photon. The steps are shown by consequent bullets (1), (2), (3), and (4). The indicated path represents the process leading to the steady state rate $\Gamma_{\mathrm{dark,ss}}^{\mathrm{(m)}}$, but other processes induced by the sequence in (a) also contribute.}
\label{figS:dark}
\end{figure}
%%%

%%%%
\subsubsection{Analytical estimation using effective operators}

Assuming the anharmonicity to exceed the classical drive frequency, the Hilbert space of the system can be truncated at the level of single photon excitation ($\mathcal{N}_1 = \mathcal{N}_2 = 1$). Analyzing the couplings, the relevant space then corresponds to the states $\{ |g,0,0\rangle, |e,0,0\rangle, |g,0,1\rangle, |e,0,1\rangle, |f,0,0\rangle, |f,0,1\rangle, |g,1,0\rangle \}$. The effective operator calculation can be performed similarly to Sec.\ref{sect:D}. In order to distinguish between $\kappa_2$ decay processes which project the system into the ground and excited subspaces, we split the jump operator $\hat{C}_{\kappa_2}$ into two parts using the projection operators $\mathcal{P}_{G} = |g,0,0\rangle\langle g,0,0| + |g,0,1\rangle\langle g,0,1|$ and $\mathcal{P}_{E} = |e,0,0\rangle\langle e,0,0| + |f,0,0\rangle\langle f,0,0| + |e,0,1\rangle\langle e,0,1| + |f,0,1\rangle\langle f,0,1|$. This gives two separate jumps $\hat{C}_{\kappa_2,G} = \mathcal{P}_{G} \hat{C}_{\kappa_2} \mathcal{P}_{G} = \sqrt{\kappa_2}|g,0,0\rangle\langle g,0,1|$ and $\hat{C}_{\kappa_2,E} = \mathcal{P}_{E} \hat{C}_{\kappa_2} \mathcal{P}_{E} = \sqrt{\kappa_2}(|e,0,0\rangle\langle e,0,1| + |f,0,0\rangle\langle f,0,1|)$. We note that this separation into two separate decay processes holds as long as interference between the $\hat{C}_{\kappa_2,G}$ and $\hat{C}_{\kappa_2,E}$ processes is suppressed. This is true once the energy separation between ground and excited states is large (in the rotating frame), and is typically justified due to hybridization by strong drive $\Omega$ and large cavity coupling $g_2$. However, this approximation can break down, and this may influence the results for the enhanced dark count rate in certain cases.

Finally, the system Hamiltonian for  finite anharmonicity (\ref{eqS:Hsys_A}) has to be projected onto relevant low-lying states, to give $\hat{\mathcal{H}}_{\mathrm{sys},A,\mathrm{dark}}$. Its non-Hermitian version then reads  $\hat{\mathcal{H}}_{\mathrm{sys},A,\mathrm{dark}}^{\mathrm{(NH)}} = \hat{\mathcal{H}}_{\mathrm{sys},A,\mathrm{dark}} - i\hat{C}_{\kappa_2,G}^\dagger \hat{C}_{\kappa_2,G}/2 - i\hat{C}_{\kappa_2,E}^\dagger \hat{C}_{\kappa_2,E}/2$.

\textit{Single dark counts.---} The relevant states of the system during an off-duty stage are shown in Fig. \ref{figS:dark}, together with possible dark count processes associated to coherent excitation out of the  ground state $|g,0,0\rangle$. We identify the most probable dark count process to be the one depicted in Fig. \ref{figS:dark}(a). This involves a two-photon Raman transition from the ground state $|g,0,0\rangle$ to the excited cavity state $|g,0,1\rangle$  through the detuned level $|e,0,0\rangle$, followed by the emission of a photon [steps (1), (2), and (3)].
The effective collapse operator for this  dark count process involving a $|g,0,1\rangle \rightarrow |g,0,0\rangle$ jump can be written as
\begin{equation}
\label{eqS:L_kappa2_G}
\hat{L}_{\kappa_2,G}^{\mathrm{eff}} = \hat{C}_{\kappa_2,G} \left[ \hat{\mathcal{H}}_{\mathrm{sys},A,\mathrm{dark}}^{\mathrm{(NH)}} \right]^{-1} \hat{V}_{\mathrm{dark}}^+,
\end{equation}
where the $\hat{V}_{\mathrm{dark}}^+ = \frac{\Omega}{2\sqrt{2}} |e,0,0\rangle \langle g,0,0|$ operator describes the excitation of the system from the ground state.

Notably, as the jump occurs within the ground state subspace, it is not amplified by the classical drive $\Omega$ and produces only a single emission event. Thus, we name the process a  \emph{single dark count}. Its rate is given by
\begin{equation}
\label{eqS:Gamma_dark_s_0}
\Gamma_{\mathrm{dark,ss}}^{\mathrm{(s)}} = \left| \langle g,0,0| \hat{L}_{\kappa_2,G}^{\mathrm{eff}} | g,0,0 \rangle \right|^2.
\end{equation} 
The rate (\ref{eqS:Gamma_dark_s_0}) can be straightforwardly computed by inverting the matrix corresponding to the non-Hermitian Hamiltonian, but its full form is too bulky to be presented here. We thus write the result of Eq. (\ref{eqS:Gamma_dark_s_0}) as a function $\Gamma_{\mathrm{dark,ss}}^{\mathrm{(s)}} (g_2,\Omega,A,\kappa_2)$ in the following. Its lowest order expansion in powers of $A^{-1}$ reads:
\begin{equation}
\label{eqS:Gamma_dark_s}
\Gamma_{\mathrm{dark,ss}}^{\mathrm{(s)}} \approx \frac{g_2^2 \Omega^2}{4 A^2 \kappa_2},
\end{equation}
showing a quadratic drop of the single dark counts with the anharmonicity of the artificial atom.

The single dark counts will be a noise source for the operation of the SPT. Fortunately, for reasonably low single count rate and large gain of the system ($N_{\mathrm{out},2}>50$), it will be possible to discriminate these single emission events coming from the classical drive from the avalanche triggered by the source photon. By setting a threshold of a certain minimum signal, the effect of the single dark counts can be eliminated. Alternatively, these single dark counts represent a leakage of the classical drive through the artificial atom. This gives rise to a weak coherent state in the output, which can be removed by interfering them with another coherent state or by offsetting the result of a final heterodyne detection (see the discussion below).

\textit{Enhanced dark counts.---} A much more severe source of noise are \emph{enhanced} or \emph{multiple dark count} where the classical drive induces a transition from the ground state to the excited manifold and triggers an avalanche of photons which is indistinguishable from the avalanche induced by a single incoming photon. We therefore search for the leading order  processes  in $A^{-1}$ which leads to the projection of the system into one of the excited states. One such process  is shown in Fig. \ref{figS:dark}(b) and represents a Raman transition to the $|g,0,1\rangle$ state [steps (1) and (2)], followed by an off-resonant excitation to $|e,0,1\rangle$ by the classical drive [step (3)], and subsequent decay to the excited state $|e,0,0\rangle$ [step (4)]. Alternatively, the system may arrive in $|f,0,0\rangle$ through a similar process.

%The major difference from the previously considered single dark count rate is the final state which should now correspond to one of the excited levels. In the SPT operational regime this corresponds to an enhancement of the parasitic signal, falsifying the arrival of the  signal we want to detect. Thus, we call it an \emph{enhanced} or \emph{multiple dark count rate}, which should be largely suppressed on the time scale relevant to experiment.

The calculation of the enhanced dark count rate proceeds in the same fashion as above, except that the final state is now an excited states so that the effective operator is
\begin{equation}
\label{eqS:L_kappa2_E}
\hat{L}_{\kappa_2,E}^{\mathrm{eff}} = \hat{C}_{\kappa_2,E} \left[ \hat{\mathcal{H}}_{\mathrm{sys},A,\mathrm{dark}}^{\mathrm{(NH)}} \right]^{-1} \hat{V}_{\mathrm{dark}}^+.
\end{equation}
The associated steady state rate is
\begin{equation}
\label{eqS:Gamma_dark_m_0}
\Gamma_{\mathrm{dark,ss}}^{\mathrm{(m)}} = \left| \langle e,0,0| \hat{L}_{\kappa_2,E}^{\mathrm{eff}} | g,0,0 \rangle \right|^2 + \left| \langle f,0,0| \hat{L}_{\kappa_2,E}^{\mathrm{eff}} | g,0,0 \rangle \right|^2,
\end{equation} 
%
%being the full formula for the enhanced rate $\Gamma_{\mathrm{dark,ss}}^{\mathrm{(m)}} (g_2,\Omega,A,\kappa_2)$. 
Again we only give the approximate expression to lowest order in $A^{-1}$ which is
\begin{equation}
\label{eqS:Gamma_dark_m}
\Gamma_{\mathrm{dark,ss}}^{\mathrm{(m)}} \approx \frac{g_2^2 \Omega^4}{32 A^4 \kappa_2}.
\end{equation}
Note that this shows a much more favorable quartic decay of the  enhanced dark counts with the anharmonicity of the artificial atom. Since $A$ is the largest scale in the problem this mean that the rate will be much lower than the single dark count rate. The rapid decay of the enhanced dark count rate is a major advantage for the SPT since it reduces the most severe type of noise in the system.

%%%%%
\subsubsection{Wave function Monte-Carlo calculation}

\begin{figure}[t!]
\includegraphics[width=1.\linewidth]{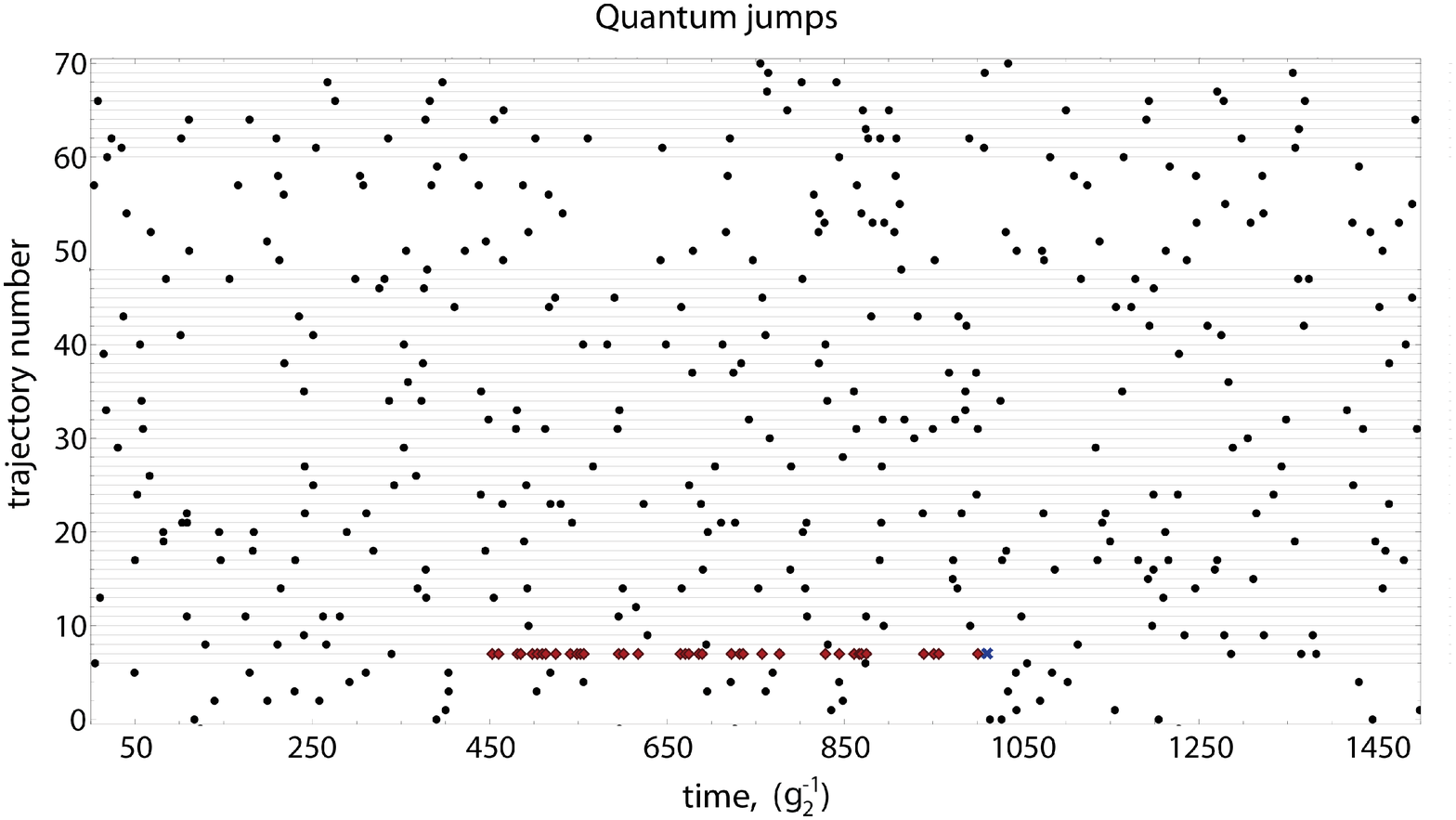}
\caption{Sample of wave function Monte-Carlo trajectories, showing $\hat{C}_{\kappa_2,G}$ (black bullets) and $\hat{C}_{\kappa_2,E}$ (dark red diamonds) dark count events. The blue cross corresponds to the $|g,1,0\rangle \rightarrow |g,0,0\rangle$ recovery process. The parameters are: $\Omega/g_2 = 2$, $\kappa_2/g_2 = 0.1$, $A/g_2 = 50$, $g_1/g_2 = 0.2$.}
\label{figS:counts}
\end{figure}

To access the dark count rate numerically we find the full counting statistics of the system taking the ground state  $|\Psi_{\mathrm{start}} \rangle = |g,0,0\rangle$ as the starting state in the absence of signal photons. We then simulate the system  using the  wave function Monte-Carlo technique (see Sec.\ref{sect:F} and Refs. \cite{Dalibard1992SM,Molmer1993SM}).  The relevant information about the dark counts is then given by the jump events provided by the ground $\hat{C}_{\kappa_2,G}$ and excited $\hat{C}_{\kappa_2,E}$ state collapse operators. To illustrate the dynamics, we show a sample of quantum trajectories, where $\hat{a}_{\mathrm{out},2}$ is monitored [Fig. \ref{figS:counts}]. Here, black dot bullets correspond to $\hat{C}_{\kappa_2,G} = \sqrt{\kappa_2}|g,0,0\rangle\langle g,0,1|$ jump events, and dark red bullets depict $\hat{C}_{\kappa_2,E}$ jumps.
%%%
%%%

Following the analysis from the previous subsection, we note the appearance of scattered single dark counts (black), with rare appearances of many-photon dark counts (dark red) coming in groups. To calculate the rates, we perform $N_{\mathrm{traj}} = 2000$ calculations each  of duration $t_{\mathrm{traj}} = 10^4 g_2^{-1}$. For the enhanced dark count rate we analyze the data given by the first jump instance in a series by making a histogram of the time intervals between jumps $\tau^{\mathrm{(m)}}$ and setting the rate to $\Gamma_{\mathrm{dark}}^{\mathrm{(m)}}= 1/ \langle \tau^{\mathrm{(m)}} \rangle$. The statistical error of the estimate is given by standard error for the $\tau^{\mathrm{(m)}}$ distribution divided by $\sqrt{N^{\mathrm{(m)}}}$, where $N^{\mathrm{(m)}}$ is total number of instances. This rate estimation procedure is only valid for recovery times (total time spent in the excited subspace) much smaller than the  total sampling time $T_{\mathrm{rec}} \ll t_{\mathrm{traj}} N_{\mathrm{traj}}$, a condition  satisfied for the chosen simulation parameters.

The results of the numerical simulations are plotted with error bars in Fig. \ref{figS:rates}. Here we have fixed the parameters to $\Omega/g_2 = 2$, $\kappa_2/g_2 = 0.1$, $g_1/g_2 = 0.2$, and consider a tunable anharmonicity $A$. The single dark count rate $\Gamma_{\mathrm{dark}}^{\mathrm{(s)}}$ is shown in Fig. \ref{figS:rates}(a). We observe a good correspondence with the full analytical estimate provided by effective operator technique [Eq. (\ref{eqS:Gamma_dark_s_0}), black solid curve], while the lowest order $A^{-2}$ expansion given by Eq. (\ref{eqS:Gamma_dark_s}) [black dashed curve] is suitable in the region of large anharmonicity.

In Fig. \ref{figS:rates}(b) we first plot the enhanced rate calculated by the wave function Monte-Carlo method (dark red error bars), and compare it to the analytical estimates given by solid (full) and dashed (simple) curves. We observe that while the full $\Gamma_{\mathrm{dark,ss}}^{\mathrm{(m)}} (g_2,\Omega,A,\kappa_2)$ function given by the solid curve [Eq. (\ref{eqS:Gamma_dark_m_0})] bares the same behavior, it gives three-to-four times smaller contribution than the wfMC calculation. This discrepancy can be attributed to the fact that effective operator formalism captures the effective rate of decay process, relevant for the steady state. However, it does not account for the temporal dynamics associated with (Rabi) oscillations between the levels, which occur after a single dark count. To examine this additional contribution, we perform an  analysis of the no-jump evolution of the system in the next subsection.

%%%%%
\subsubsection{No-jump evolution calculation}

An additional way to extract the information about rate of going outside the ground state $|g,0,0\rangle$ is to calculate the no-jump  evolution \cite{Molmer1993SM}, where the norm of system's wave function $|\psi(t)\rangle$ decays under the propagation  with the non-Hermitian Hamiltonian $\hat{\mathcal{H}}_{\mathrm{sys},A,\mathrm{dark}}^{\mathrm{(NH)}}$. The jump rate is then proportional to $\Gamma_j \propto \delta p/\delta t = \langle \psi(t)|\hat{C}_{j}^\dagger \hat{C}_{j}|\psi(t)\rangle $, where $\hat{C}_{j}$ is the collapse operator for process $j$. To get the actual rate the change of a particular jump probability shall be divided by the total norm, $\Gamma_j (t) = \langle \psi(t)|\hat{C}_{j}^\dagger \hat{C}_{j}|\psi(t)\rangle / \langle \psi(t)|\psi(t)\rangle $. The rates calculated by the effective operators correspond to those obtained by adiabatic elimination and represent the steady state rates $\Gamma_{j,\mathrm{ss}} = \Gamma_j (t\rightarrow \infty)$. For instance, the single dark count rate $\Gamma_{\mathrm{dark,ss}}^{\mathrm{(s)}}$ can be extracted from $\langle \psi(t)|\hat{C}_{\kappa_2,G}^\dagger \hat{C}_{\kappa_2,G}|\psi(t)\rangle / \langle \psi(t)|\psi(t)\rangle $ at large $t$, and the result of this coincides with the full wfMC calculation results (not shown).

\begin{figure}[t!]
\includegraphics[width=1.\linewidth]{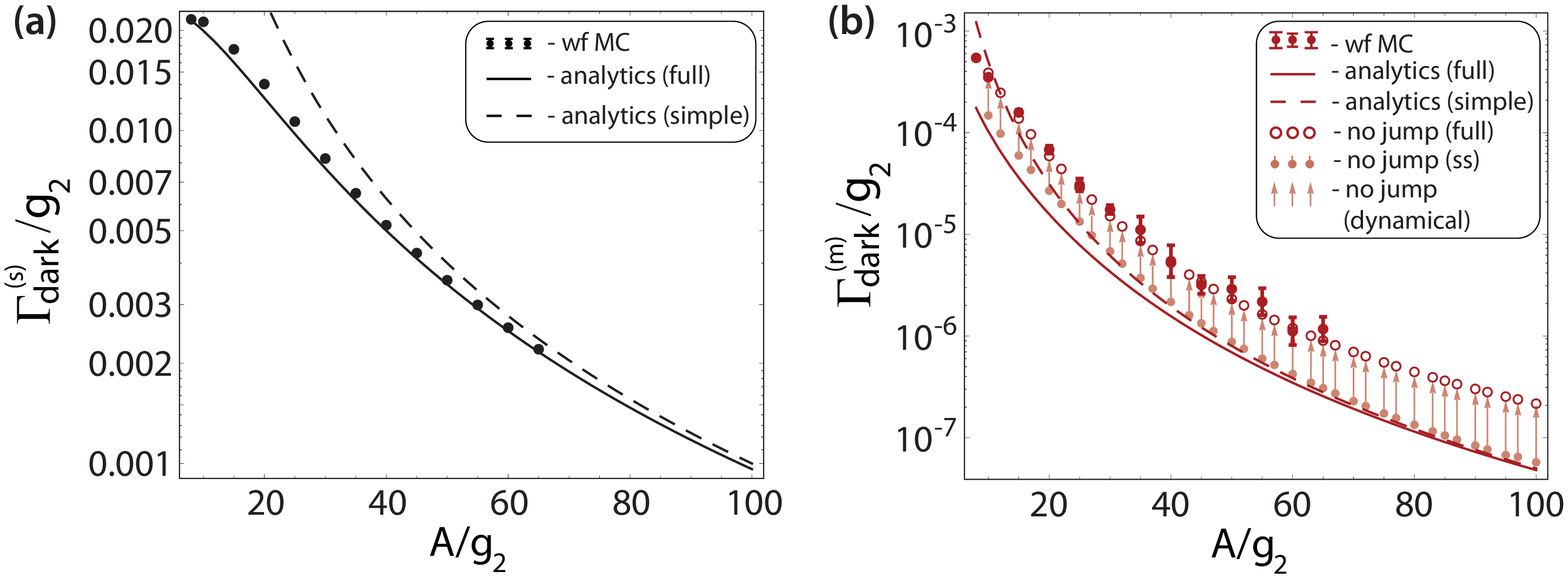}
\caption{Dark count rates. (a) Single dark count rate plotted as a function of anharmonicity $A$. The dots correspond to the results of wave function Monte-Carlo simulation. These results are compared to the full analytical solution in Eq.  (\ref{eqS:Gamma_dark_s_0}). The simplified formula for $\Gamma_{\mathrm{dark,ss}}^{\mathrm{(s)}}$ is shown by the dashed curve. (b) Multiple dark count rates calculated using different procedures. Solid and dashed curves represent analytical results provided by Eqs. (\ref{eqS:Gamma_dark_m_0}) and (\ref{eqS:Gamma_dark_m}), respectively. Vertical arrows show the the enhancement  from the steady state value (dots) to the total dark count rate (circles) due to dynamical contributions induced by the single dark count. The total dark count rate is calculated both from a full Monte Carlo simulation (filled circles) as well as from the no-jump evolution following a single dark count (open circles). The results of the wave function Monte Carlo simulation include error bars, but these are hardly visible for most of the points. 
The parameters are: $\Omega/g_2 = 2$, $\kappa_2/g_2 = 0.1$, $g_1/g_2 = 0.2$, $N_{\mathrm{traj}} = 2000$, $t_{\mathrm{traj}} = 10^4 g_2^{-1}$.}
\label{figS:rates}
\end{figure}

The estimate of the enhanced dark count rate relies on the jump rate for going to the excited state subspace. Compared to the single photon dark count process, however, the steady state rate $\Gamma_{\mathrm{dark,ss}}^{\mathrm{(m)}}$ is much lower for the multiple dark counts, and the influence of higher order effects are therefore important. In particular, the presence of a small coherent coupling between the ground and excited states leads to oscillations in the probability to jump to the excited states $P_{\mathrm{E}}^{(\mathrm{jump})}(t)$ following a single dark count. We find that these oscillations tend to enhance the jump probability $P_{\mathrm{E}}^{(\mathrm{jump})}$ during the time span where a jump is more probable. This means that the single dark counts provide an induced rate of multiple dark counts. Although this probability is small, the rate of the single dark counts is much higher. Hence the single dark counts induce an additional \emph{dynamical} enhanced rate contribution which is comparable to the steady state value found from the effective operators. The total enhanced dark count rate reads $\Gamma_{\mathrm{dark}}^{\mathrm{(m)}}(t_{\mathrm{end}}) = \left( \int\limits_{t_0}^{t_{\mathrm{end}}} \langle \psi(t)|\hat{C}_{\kappa_2,E}^\dagger \hat{C}_{\kappa_2,E}|\psi(t)\rangle dt \right) / \left( \int\limits_{t_0}^{t_{\mathrm{end}}} \langle \psi(t)|\psi(t)\rangle dt \right)$, where we assigned $t_0$ as a starting time representing the time of a single dark count and $t_{\mathrm{end}}$ denotes an integration time. This allows to capture the increased probability to jump to excited states, provided by oscillations due to level dressing. While the expression for multiple count rate $\Gamma_{\mathrm{dark}}^{\mathrm{(m)}}(t_{\mathrm{end}}\rightarrow +\infty)$ contains full information about dark counts, its convenient to separate it into a dynamical contribution, $\Gamma_{\mathrm{dark,dyn}}^{\mathrm{(m)}}= \Gamma_{\mathrm{dark}}^{\mathrm{(m)}}(T)$, where $T$ is a short period at which oscillations occur, and the subsequent steady state contribution. The latter can also be defined as $\Gamma_{\mathrm{dark,ss}}^{\mathrm{(m)}} = \lim\limits_{t\rightarrow +\infty} \langle \psi(t)|\hat{C}_{\kappa_2,E}^\dagger \hat{C}_{\kappa_2,E}|\psi(t)\rangle / \langle \psi(t)|\psi(t)\rangle$.
The results of full no-jump calculations are presented in Fig. \ref{figS:rates}(b) by red circles, whereas the full dots denote the steady state rate contribution, well approximated by the analytically derived effective rate.
%%%
%%%

The previously described procedure provides an efficient numerical estimation of the induced rate, but to gain more insight into the nature of this we also consider a simple analytical description of the causes of the dynamical jump contribution. To this end we perform the analysis in a suitable dressed state picture. Namely, during the single dark count cycle [Fig. \ref{figS:dark}(a)] which corresponds to an effective $|g,0,0 \rangle$ dephasing  at a rate $\Gamma_{\mathrm{dark}}^{\mathrm{(s)}}$, the jump processes should not be to the original states, but should include the weak mixing of the ground state with the excited states. Here, the ground state contains a weak admixture of excited state, $|g,0,0\rangle \mapsto \widetilde{|g,0,0\rangle} = |g,0,0\rangle + \frac{\Omega}{2\sqrt{2}}|e,0,0\rangle$. Similarly, the excited level gets a ground state contribution, $|e,0,0\rangle \mapsto \widetilde{|e,0,0\rangle} = |e,0,0\rangle - \frac{\Omega}{2\sqrt{2}}|g,0,0\rangle$. This means that each single dark count event can induce an additional multi-dark count. The process considered here corresponds to the collapse operator $\hat{C}_{\Gamma_{\mathrm{dark}}^{\mathrm{(s)}}} = \sqrt{\Gamma_{\mathrm{dark}}^{\mathrm{(s)}}} |g,0,0\rangle \langle g,0,0|$, which describes associated dephasing of the ground state. From this we can find the relevant rate of accidental jumps to the excited state during a single dark count cycle,
\begin{equation}
\label{eqS:Gamma_s_dyn}
\Gamma_{\mathrm{dark,dyn}}^{\mathrm{(m)}} = \left| \widetilde{\langle e,0,0|} \hat{C}_{\Gamma_{\mathrm{dark}}^{\mathrm{(s)}}} \widetilde{|g,0,0 \rangle} \right|^2 = \frac{g_2^2 \Omega^4}{32 A^4 \kappa_2},
\end{equation}
which gives a dynamical contribution equal to the steady state rate described by Eq. (\ref{eqS:Gamma_dark_m}), thus providing an analytical multi-dark count rate estimate closer to the numerical simulation. Furthermore, we find that additional dynamical channels depend on the $|g,0,0\rangle \longleftrightarrow |g,1,0\rangle \longleftrightarrow |e,0,1\rangle$ coupling, suggesting that processess other than the direct $|g,0,0\rangle \longleftrightarrow |e,0,0\rangle$ process shall be considered. Regardless of the exact cause of the dynamical contributions, however, these effects will be a perturbation on the single dark count rate. Therefore the multiple dark count rate will be much lower than the single dark count rate, thus providing an efficient suppression  of the most detrimental source of noise. 

The full description of dynamical dark count processes involves a coupled system of six relevant levels, and in principle shall also include the interference effects between different decay paths, neglected due to the separation of the collapse operator $\hat{C}_{\kappa_2}$. While the latter can potentially change the dark count rate, so that constructive interference could enhance it by a factor of two, or destructive interference could diminish the rate, a full analytical treatment is complicated and a more detailed analysis is beyond the scope of this paper.

%%%%%%%%%%%%%%%%%%%%%%%%%%%%%%%%%%%%%%%%%%%%%%%

\subsection{\label{sect:I}Single photon detection}

In the main text of the paper we describe a reliable way to create the single photon transistor device. It generates an avalanche of gain photons conditioned on the arrival of a single signal photon. Consequently, this process can be used as a single microwave photon detector. Here we describe a possible procedure to achieve efficient detection of the outgoing signal using existing homodyne detection techniques.
%%%
\begin{figure}[t!]
\includegraphics[width=0.75\linewidth]{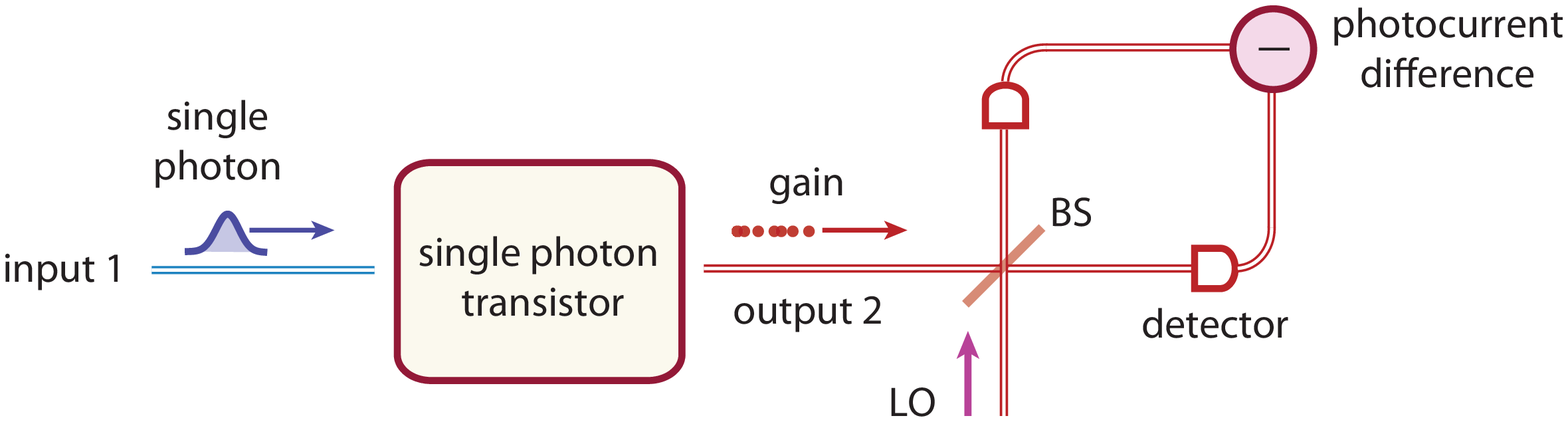}
\caption{Microwave photon detection scheme. The signal photon enters a single photon transistor device, which generates an avalanche of gain photons at the output. This output signal is mixed on the beam-splitter (BS) with a strong coherent drive of a microwave local oscillator (LO). Following the detection of microwave photons at each arm, the photocurrent difference signal can be extracted.}
\label{figS:detection}
\end{figure}
%%%

The measurement of a signal photon relies on the possibility to detect a large number of gain photons, distributed on different frequency modes. This can conveniently be done using a heterodyne detection scheme \cite{WallsMilburnSM,Kimble1995SM,Walmsley2000SM}. The corresponding process requires the mixing of the outgoing gain photons with an additional coherent microwave drive, acting as a local oscillator (LO) at a frequency $\omega_0$. The detection of gain photon then depends on the measurement of the difference photocurrent [see sketch in Fig. \ref{figS:detection}]. It is generically given by the relation
\begin{equation}
\label{eqS:I_diff}
\hat{I}_{-} = \frac{1}{2} \left[\alpha(t) + \hat{a}_{\mathrm{gain}}(t)\right]^\dagger \left[\alpha(t) + \hat{a}_{\mathrm{gain}}(t)\right] - \frac{1}{2} \left[\alpha(t) - \hat{a}_{\mathrm{gain}}(t)\right]^\dagger \left[\alpha(t) - \hat{a}_{\mathrm{gain}}(t)\right] = \alpha^*(t) \hat{a}_{\mathrm{gain}}(t) + \alpha(t) \hat{a}_{\mathrm{gain}}^\dagger(t),
\end{equation}
where $\alpha(t) = \alpha \exp(-i\omega_0 t)$ is the local oscillator with amplitude $\alpha$, and $\hat{a}_{\mathrm{gain}}(t)$ describes the gain photons coming out of SPT. We note that a similar scheme can be arranged using Josephson parametric amplifiers and high-electron-mobility transistors \cite{Mallet2011SM}.

Assuming a real amplitude for the coherent state, and decomposing the SPT output into an infinite set of frequency modes, $\hat{a}_{\mathrm{gain}}(t) = \sum\limits_k^{\infty} \hat{a}_{k}\exp(-i\omega_k t)/\sqrt{T}$, we get
\begin{equation}
\label{eqS:I_diff_2}
\hat{I}_{-} = \frac{\alpha}{\sqrt{T}} \sum\limits_k^{\infty} \left(\hat{a}_{k}e^{-i(\omega_k - \omega_0) t} + \hat{a}_{k}^\dagger e^{i (\omega_k - \omega_0) t} \right),
\end{equation}
where $\omega_k = 2\pi k/T$ denotes the frequency of each mode and $T$ is the integration time. While in general the frequency range spans to infinity, the outgoing photons will be distributed in a bandwidth $\delta \omega$ set by the qubit dynamics. The modes in the relevant frequency range can be isolated by performing a Fourier transform of the output current and introducing a suitable filter function. For the discrete Fourier transform considered here this changes the summation range to a finite number. Finally, the Fourier transform of the difference signal reads $\hat{I}_{-} = \alpha \sum\limits_{k=1}^{M} \left(\hat{a}_{k} + \hat{a}_{-k}^\dagger \right)$, where indices $\pm k$ correspond to different sidebands. While the number of frequency modes at the output $M$ can be chosen at will, the relevant frequency range in which the gain photons come out can be easily estimated for the relevant SPT parameters. In particular, the frequency spacing between modes of the output is $\delta\omega \sim \sqrt{g_2^2 + \Omega^2}$, and the frequency integration window is proportional to $1/T = \gamma_{\mathrm{rec}}$, where $\gamma_{\mathrm{rec}}$ is the recovery rate. For the cavity 2 decay rate $\kappa_2$ being comparable to the coupling constants $g_2$ and $\Omega$, in the worst case the photon spectrum will fill the entire frequency width $\delta \omega$. Their total number can be then estimated by $M=\delta\omega T/2\pi$. In the opposite case of small outcoupling rate $\kappa_2$ this relation will be modified, leading the much smaller number of modes as compared to the number of gain photons. However, this also comes with a smaller gain [see Fig. 3(c) in the main text].

We proceed with estimating the signal-to-noise ratio for the detection. For this, we assume that the experiment extracts the information by squaring the output from each mode and summing over all modes (note that one could imagine more advanced statistical analysis of the output \cite{Walmsley2000SM}, but for simplicity we restrict ourselves to this strategy). This is described by the observable operator $\hat{O} = \alpha^2\sum\limits_{k=1}^M (\hat{a}_k^\dagger + \hat{a}_{-k}) (\hat{a}_k + \hat{a}_{-k}^\dagger)$, and we also introduce the square of this operator, $\hat{O}^2 = \alpha^4 \sum\limits_{k=1}^M \sum\limits_{k'=1}^M (\hat{a}_k^\dagger + \hat{a}_{-k}) (\hat{a}_k + \hat{a}_{-k}^\dagger) (\hat{a}_k'^\dagger + \hat{a}_{-k'}) (\hat{a}_k' + \hat{a}_{-k'}^\dagger)$. Next, we need to consider expectation values of these operators over the vacuum state $|\mathrm{vac}\rangle$ and SPT output given by $|\mathrm{out}\rangle$. The signal-to-noise ratio analysis requires us to determine the mean and the variance of photon number distributions for vacuum (dark counts) and signal input (gain photons).  

A vacuum input state corresponds to an output with zero photons $|\mathrm{vac}\rangle = |0_1 0_2 .. 0_M\rangle$ written in the Fock states basis. We note that the single dark counts can be understood as the leakage of the drive $\Omega$ to the output 2 port. This results in one of the output modes being in a coherent state. The resulting change can be accounted for by adding a displaced coherent state to one of the input modes. We consider the vacuum case for simplicity, but the overall result is expected to be the same once the constant shift due to the non-zero coherent state amplitude is subtracted from the particular mode in question.
The estimate of the mean value of the observable for the vacuum input is given by $\langle \mathrm{vac} | \hat{O} | \mathrm{vac} \rangle = \langle \mathrm{vac} | \sum\limits_{k=1}^M \hat{a}_{-k}\hat{a}_{-k}^\dagger | \mathrm{vac} \rangle = M $ (we set $\alpha=1$ for simplicity). The expectation value of $\langle \hat{O}^2 \rangle_{\mathrm{vac}}$ is represented by two non-zero terms $\langle \mathrm{vac} | \sum\limits_k^M \sum\limits_{k'}^M \hat{a}_{-k}\hat{a}_{k}\hat{a}_{k'}^\dagger\hat{a}_{-k'}^\dagger | \mathrm{vac} \rangle = M$ and $\langle \mathrm{vac} | \sum\limits_k^M \sum\limits_{k'}^M \hat{a}_{-k}\hat{a}_{-k}^\dagger\hat{a}_{-k'}\hat{a}_{-k'}^\dagger | \mathrm{vac} \rangle = M^2$. This yields the variance of $\mathrm{Var}[\hat{O}] = \langle \hat{O}^2 \rangle_{\mathrm{vac}} - \langle \hat{O} \rangle_{\mathrm{vac}}^2 = M$. From the central limit theorem it follows that since a large number of the modes $M$ is considered, the distribution will be Gaussian, $P_{\mathrm{vac}}(O) = \exp[-(O-M)^2/2M]/\sqrt{2\pi M}$.

Next, we proceed with the description of the difference photon number distribution for the case of a general input. This can be expressed in the Fock state basis as an arbitrary superposition of the states $|\mathrm{out}\rangle = \sum\limits_j C_j |n_{1,j} n_{2,j} .. n_{M,j} \rangle$. To obtain an estimate, we assume a non-squeezed output state, $\langle \mathrm{vac} | \sum\limits_k^M \hat{a}_{k}\hat{a}_{k} | \mathrm{vac} \rangle = 0$. The mean value then reads
\begin{equation}
\label{eqS:O_out_mean}
\langle \hat{O} \rangle_{\mathrm{out}} = \langle \mathrm{out} | \sum\limits_k^M \hat{a}_{k}^\dagger \hat{a}_{k} | \mathrm{out} \rangle + \langle \mathrm{out} | \sum\limits_k^M \hat{a}_{-k}\hat{a}_{-k}^\dagger | \mathrm{out} \rangle = 2 N + M,
\end{equation}
where $N$ denotes the total number of photons at the output. Thus the efficient operation of the system as a detector requires $N\gg \sqrt{M/4}$, which typically can be achieved for large gain parameters.

To provide the quantitative estimates the statistics of the output has to be considered carefully. In general this is an arduous task as the numerics only allows us to access the probability distribution for the region of small gain, which does not fall into parameter regime we are generally interested in. Instead we shall estimate the outcome of this using the inferred probability distribution from the simulation in Sec. F. To be concrete we then set a threshold $O_{\mathrm{th}}$ for faithful detection such that a detection event is registered if $O > O_{\mathrm{th}}$, whereas no detection is observed if $O < O_{\mathrm{th}}$. This implies that the process of distinguishing the signal from vacuum introduces two types of errors: a reduction of the efficiency from output signal not exceeding the threshold, and a dark count rate from vacuum fluctuations exceeding the threshold. The trade-off between these to quantities is ultimately set by $O_{\mathrm{th}}$ and changes with system properties, such as gain, recovery rate, and the number of frequency modes.

%First, we consider the counting statistics of the output signal analyzed in details in Section \ref{sect:F}. This can be approximated by the two-parameter gamma distribution given by $P_{\mathrm{out}}^{\Gamma}(O) = \frac{1}{\lambda^\xi \Gamma(\xi)} O^{\xi - 1} \exp(-O/\lambda)$, as can be seen from Figs. \ref{figS:Pn} and \ref{figS:prob_large}. Here the parameter $\xi$ defines the shape of probability distribution at small $O$, and the corresponding mean value reads $\langle \hat{O} \rangle_{\mathrm{out}}^{\Gamma} = \xi \lambda$, which relates $\lambda$ to the number of gain photons. The estimate of the efficiency $\eta$ of the detection then relies on the integration of the probability distribution $P_{\mathrm{out}}^{\Gamma}(O)$ in the interval $[O_{\mathrm{th}},\infty]$. Taking the parameters of $\Omega = g_2/2$, $g_1/g_2 = 0.003$, $\kappa_2/g_2 = 1$, and the corresponding gain of 200 photons, we can estimate the number of modes $M \approx 90$. The distribution parameter is fitted by $\xi = 3$, and we choose the threshold to be in the form $O_{\mathrm{th}} = \zeta \sqrt{M}$, where $\zeta$ is a tunable parameter. Complementary, the estimate for dark counts is given by the integral of $P_{\mathrm{vac}}(O)$ in the relevant range of photon numbers $[O_{\mathrm{th}},\infty]$. Considering $\zeta = 4$ we get $\eta = 99.7\%$ efficiency with $3\times 10^{-5}$ dark count probability. Furthermore, if we account for the increased probability to terminate the gain cycle at the smaller number of counts (fitted from \ref{figS:prob_large}), the efficiency drops to $\eta = 99.4\%$.

First, we consider the worst case scenario where the output is fully determined by the exponential distribution given as $P_{\mathrm{out}}^{\exp}(O) = \exp(-O/\lambda)/\lambda$, with a mean value of $\langle \hat{O} \rangle_{\mathrm{out}}^{\exp} = \lambda$. Taking the parameters of $\Omega = g_2/2$, $g_1/g_2 = 0.003$, $\kappa_2/g_2 = 1$, and the corresponding gain of $200$ photons, we can estimate the number of modes $M \approx 90$. We choose the threshold to be in the form $O_{\mathrm{th}} = \zeta \sqrt{M}$, where $\zeta$ is a tunable parameter. Setting it to $\zeta = 2$ we estimate an efficiency of $\eta = 95\%$ with $0.02$ dark count probability. Given the deviation of gain photon probability distribution from the exponential at the small counts region, we expect the results to be slightly improved for the actual detection. Alternatively, we can modify the parameters to achieve higher gain of 1000 photons, choosing $\Omega = g_2/2$, $g_1/g_2 = 0.0013$, $\kappa_2/g_2 = 1$, and estimating $M = 600$ as a number of frequency modes. By setting $\zeta = 3$ we get an efficiency $\eta = 97\%$ at a $10^{-3}$ dark count level.

Next, we note that one can consider the case of an undercoupled second cavity, $\kappa_2/g_2 \ll 1$, where number of frequency modes of the output decreases. This corresponds to two frequency bands determined by the allowed transitions between dressed states shown in Fig. 1(c) of the main text. Then, the gain can be detected by selecting two frequency bands containing the photons of interest from the output of the heterodyne detection. The described scheme can potentially reduce the effects of vacuum noise, as compared to multimode detection.

Finally, we note that exploiting the single photo transistor, the possible detection scheme is not restricted to the optics-inspired techniques. In particular, using a calorimetric device the measurement of zeptojoule microwave pulses is possible \cite{Govenius2016SM}. In our setup this corresponds to a signal of 130 photons at a frequency of $12$ GHz, thus opening the possibility of temperature based detection of a single photon.

%%%%%%%%%%%%%%%%%%%%%%%%%%%%%%%%%%%%%%

\end{document}